\documentclass[pdflatex,sn-mathphys-num,iicol]{sn-jnl}

\usepackage{graphicx}%
\usepackage{multirow}%
\usepackage{amsmath,amssymb,amsfonts}%
\usepackage{amsthm}%
\usepackage{mathrsfs}%
\usepackage[title]{appendix}%
\usepackage{xcolor}%
\usepackage{textcomp}%
\usepackage{manyfoot}%
\usepackage{booktabs}%
\usepackage{listings}%
\usepackage{threeparttable}

\newcommand{\crpropa}{\texttt{CRPropa3.2.1}}
\begin{document}

\title[Modeling individual nearby radio galaxies as ultra-high-energy cosmic-ray accelerators]{Modeling individual nearby radio galaxies as ultra-high-energy cosmic-ray accelerators}


\author*[1]{\fnm{Andressa} \sur{Colaço}}\email{andressacolaco@usp.br}
\author[1]{\fnm{Gabriel} \sur{Azeredo}}\email{gabriel@ifsc.usp.br}
\author[1]{\fnm{Isadora} \sur{Parillo}}\email{isadora.parillo@usp.br}
\author[2]{\fnm{Cainã} \spfx{de} \sur{Oliveira}}\email{caina.oliveira@usp.br}
\author[1]{\fnm{Vitor} \spfx{de} \sur{Souza}}\email{vitor@ifsc.usp.br}

\affil[1]{\orgdiv{Instituto de Física de São Carlos}, \orgname{Universidade de São Paulo}, \orgaddress{\street{Av. Trabalhador São Carlense, 400}, \city{São Carlos}, \postcode{13566-590}, \state{São Paulo}, \country{Brazil}}}

\affil[2]{\orgdiv{Instituto de Física}, \orgname{Universidade de São Paulo}, \orgaddress{\street{Rua Do Matão 1371}, \city{São Paulo}, \postcode{05508-090}, \state{São Paulo}, \country{Brazil}}}


\abstract{Nearby radio galaxies are among the most promising candidates for the acceleration of ultra-high-energy cosmic rays (UHECRs). In this work, we develop a physically motivated, source-resolved framework to quantify the contribution of the three nearest Fanaroff-Riley Class I (FR-I) radio galaxies --- Centaurus~A, Virgo~A, and Fornax~A --- to the UHECR flux measured by the Pierre Auger Observatory. Acceleration spectra derived from detailed jet-acceleration models are combined with numerical simulations of extragalactic propagation, while the more distant radio-galaxy population is treated as a continuous background. By fitting our simulation exclusively to the measured UHECR energy spectrum, we determine the relative contribution of each source and constrain the fraction of jet power converted into UHECR luminosity. We find that a small number of nearby radio galaxies can account for the highest-energy UHECR flux with acceleration efficiencies of order $10^{-3}$-$10^{-2}$, while the background contribution remains subdominant. Some resulting scenarios yield mass-composition trends broadly consistent with observations at the highest energies and predict distinct levels of secondary neutrino fluxes. These results demonstrate that physically grounded, source-specific modeling of nearby radio galaxies provides a physically motivated contribution to the interpretation of the highest-energy UHECR component.}


\maketitle

\section{Introduction}
\label{sec:intro}

Ultra-high-energy cosmic rays (UHECRs), with energies exceeding $\sim 10^{18}\,\mathrm{eV}$, remain among the most challenging open problems in astrophysics. Although extensive measurements by the Pierre Auger Observatory and the Telescope Array have established the global features of the energy spectrum, mass composition, and arrival-direction distributions, the identification of the sources responsible for the highest-energy particles is still unresolved. Any viable model must simultaneously account for the observed spectral suppression at the highest energies, the energy-dependent evolution of mass-sensitive air-shower observables, and the anisotropies reported in arrival directions across angular and energy scales.

Radio galaxies have long been recognized as promising UHECR source candidates, as their relativistic jets and extended radio lobes provide large acceleration volumes, long activity timescales, and substantial kinetic power~\cite{2022-rieger}. In particular, nearby FR-I radio galaxies combine favorable energetics with short propagation distances, thereby minimizing energy losses and making their potential contribution directly testable with current UHECR data. Among them, Centaurus~A (Cen~A), Virgo~A (Vir~A), and Fornax~A (For~A) stand out as the most prominent local systems, and they have been repeatedly discussed as benchmark sources in both theoretical and phenomenological studies.

Most previous investigations of radio galaxies as UHECR sources have relied on simplified assumptions for the injected particle spectrum, typically adopting a single power-law motivated by diffusive shock acceleration and treating source populations in a largely homogeneous manner~\cite{Eichmann_2018, Eichmann_2022, de_Oliveira_2021}. In contrast, recent advances in the modeling of particle acceleration in relativistic jets indicate that multiple acceleration mechanisms --- including shock, shear, and stochastic processes --- can operate simultaneously, leading to injection spectra with non-trivial shapes that deviate from simple power laws~\cite{Seo_2023, Seo_2024, Comisso_2024, Mbarek_2019}. At the same time, growing observational evidence suggests that individual nearby sources may have undergone complex evolutionary histories, implying that source-specific properties can play a decisive role in shaping the observed UHECR flux.

The present work introduces a new, physically driven framework to test the role of nearby radio galaxies as UHECR sources. For the first time, we combine acceleration spectra derived from detailed jet-acceleration studies with source-specific properties of individual nearby FR-I radio galaxies, and we confront this framework directly with the UHECR energy spectrum measured by the Pierre Auger Observatory. Our approach departs from purely population-based treatments by modeling Cen~A, Vir~A, and For~A individually, while simultaneously accounting for the cumulative contribution of the more distant radio galaxy population through a continuous distribution of sources. This hybrid strategy allows us to capture both the distinctive characteristics of the nearest sources and the unavoidable background from the broader radio-galaxy population.

A key strength of this work is the use of a minimal set of physically motivated parameters. Rather than fitting arbitrary injection spectra or compositions, we restrict our modeling choices to scenarios grounded in current theoretical and observational understanding of AGN jets, including acceleration efficiencies tied to jet power and composition scenarios motivated by astrophysical enrichment processes. Within this constrained framework, we determine the relative contribution of each nearby source and the background population by fitting exclusively to the measured UHECR energy spectrum. The resulting best-fit solutions then yield non-trivial predictions for the mass composition at Earth and for the associated flux of secondary neutrinos, providing additional, independent tests of the model.

By linking source-specific jet properties, realistic acceleration spectra, and numerical extragalactic propagation in a unified analysis, this study represents a significant step beyond previous phenomenological treatments of radio galaxies as UHECR sources. Our results demonstrate that a small number of well-motivated nearby radio galaxies, supplemented by a subdominant background population, can reproduce key features of the observed UHECR spectrum with effective source normalizations that are broadly consistent with current expectations for the energetics of AGN jets. In doing so, this work highlights the explanatory power of source-resolved modeling and provides a physically motivated framework for future multi-messenger tests of UHECR origin scenarios.

The present work is not intended to provide a complete global description of the entire cosmic-ray spectrum across all energies and observables. Instead, we focus specifically on the highest-energy component of the UHECR flux, where nearby extragalactic sources are expected to play the dominant role due to propagation horizons and energy losses. Consequently, the energy ranges adopted in the fits are restricted to energies above 5 or 20 EeV, depending on the scenario considered. The primary limitation of this work is that it neglects the effect of magnetic fields in the cosmic-ray propagation. The present study intentionally focuses on the source energetics and spectral viability of nearby radio galaxies under simplified propagation assumptions. A full treatment of magnetic-field effects requires three-dimensional simulations and introduces substantial uncertainties associated with the poorly constrained structure of the EGMF and GMF.

This paper is organized as follows. In Section~\ref{sec:rgs} we describe the source emission models, including the adopted acceleration spectra, composition scenarios, and the treatment of individual and background radio galaxies. Section~\ref{sec:methods} details the extragalactic propagation setup and the statistical procedure used to fit the UHECR energy spectrum. The resulting source contributions, composition predictions, and neutrino fluxes are presented in Section~\ref{sec:results}. We discuss the physical implications of our findings in Section~\ref{sec:discussion}, and we summarize our conclusions in Section~\ref{sec:conclusion}.

\section{Radio galaxies as UHECR sources}
\label{sec:rgs}

In this section, we describe the properties adopted for radio galaxies as potential sources of UHECRs. Our implementation considers details of: a) the energy spectrum, b) the mass composition, c) the emitted power, and d) the source distribution in the Universe. We follow a two-pronged approach: the nearest and brightest sources are modeled individually (Centaurus A, Fornax A and Virgo A), allowing us to account for their specific physical characteristics, while the remaining, more distant population of radio galaxies is treated statistically through a continuous distribution.
\subsection{Energy spectrum and composition}
\label{sec:acceleration_models}

We use two functional forms to describe the emitted energy spectrum of radio galaxies: a) power-law with exponential cutoff (PLEC) and b) a double power-law with exponential cutoff (DPLEC).

Canonical Fermi acceleration and subsequent developments~\citep{1949_Fermi, Bell:1978zc, PROTHEROE1999} suggest that diffusive shock acceleration (DSA) results in an emitted energy spectrum given by a power-law suppressed by an exponential cutoff
\begin{equation} 
\label{eq:fermi}
\left(\frac{dN}{dE}\right)_{\text{PLEC}} = \bigg(\frac{E}{ E_{\text{break}}}\bigg)^{-s} \exp{\left(-\frac{E}{ E_{\text{break}}}\right)}
\end{equation}
where the spectral index $s$ is approximately $s = 2$ for non-relativistic shocks and $s \approx 2.2-2.3$ for relativistic shocks~\cite{2002_Lemoine}. The exponential cutoff accounts for the finite size and/or the finite acceleration power of the source. 

The calculations performed by Seo et al.~\cite{Seo_2024} for FR-I radio galaxies considering multiple acceleration mechanisms such as DSA, turbulence, and shear acceleration in the jet and in the environment resulted in emitted energy spectrum described by a double power-law with an exponential cutoff

{\small
\begin{align} 
 \left(\frac{dN}{dE}\right)_{\text{DPLEC}} = 
 \Bigg[{\bigg(\frac{E}{ E_{\text{break}}}\bigg)^{2.6} + \bigg(\frac{E}{ E_{\text{break}}}\bigg)^{0.6}} \Bigg]^{-1} \nonumber \\
 \exp{\left(-\frac{E}{ E_{\text{break}}  \Gamma_{\text{eff}}^2}\right)},\label{eq:fri_spectrum}
\end{align}
}
\noindent where $E_{\text{break}}$ is the break energy of the power-law and $\Gamma_{\rm eff}$ is the mean Lorentz factor of the spine of the jet. 

The break energy depends on the acceleration efficiency and escape, being approximately equal to the Hillas energy for efficient acceleration models~\citep{Seo_2024}. Motivated by the Hillas-Lovelace limit~\citep{1976_Lovelace, 1984_Hillas}, we estimate $E_{\text{break}}$ as a function of the jet power $Q_{\rm jet}$ as can be seen in Appendix~\ref{appendix:break_energy}). The same value of $E_{\rm break}$ can be assumed for both the PLEC and DPLEC energy spectra. 

Nongradual shear acceleration boosts the maximum energy achievable for particles by a factor $\Gamma_{\rm eff}^2$, in the so-called espresso acceleration~\citep{Caprioli_2015}. For many AGNs, $\Gamma_{\rm eff}$ and $Q_{\rm jet}$ can be estimated from electromagnetic measurements~\citep{Meyer2017-en,Godfrey2016}. Table~\ref{tab:acc_parameters} shows $Q_{\rm jet}$, $R_{\rm break}$ and $\Gamma_{\rm eff}$ for Centaurus A, Fornax A and Virgo A.
 
\begin{table*}[t]

\caption{Parameters for Centaurus A, Fornax A and Virgo A. }
\label{tab:acc_parameters}

\begin{tabular*}{\textwidth}{@{\extracolsep\fill}ccccc}
\toprule
 RG & Distance (Mpc)&$Q_{\rm jet}$ ($10^{43} {\rm erg \ s^{-1}}$) & $R_{\rm break}$ ($\rm EV$) & $\Gamma_{\rm eff} $ \\
\midrule
    Cen~A & $3.8$& 2~\cite{Neff_2015} & 5.7  &  1.2 \cite{Snios_2019}  \\
    Vir~A & $16.4$& 5  \cite{Anantua_2023}    & 8.2  &  2.5 \cite{Mertens_2016}  \\
    For~A  & $20.8$ & 0.24 \cite{Maccagni_2021} & 2.4  &  1.5 \cite{Mullin_2009}$^{a}$  \\
    \bottomrule

\end{tabular*}
\text{\small $~^{a}$ The value was estimated based on a class of sources with similar characteristics to Fornax A. \cite{Maccagni_2020}.}
\end{table*}

The composition of UHECR emitted by a source is largely unknown. We consider two astrophysically motivated hypotheses: \textit{Solar} and \textit{Wolf-Rayet}. \textit{Solar} is a light mass composition, where the elemental abundances match those of the Solar System~\citep{Eichmann_2019}. \textit{Wolf-Rayet} is an intermediate mass composition, based on the abundance of elements in Wolf-Rayet stellar winds~\citep{Liu_2012, de_Oliveira_2021, Muller_2025}. The abundance of each representative nucleus fraction is presented in Table~\ref{tab:compositiontable}.
    
\begin{table*}[t]
\caption{Fraction of elements for \textit{Solar} and \textit{Wolf-Rayet} composition hypotheses.}
\label{tab:compositiontable}

\begin{tabular*}{\textwidth}{@{\extracolsep\fill}cccccc}
\toprule
    Composition & $f_{\rm H}$ & $f_{\rm He}$ & $f_{\rm N}$ & $f_{\rm Si}$ & $f_{\rm Fe}$ \\ 
    \midrule
    Solar       & 0.922    & 0.078     & $8\times 10^{-4}$     & $8\times 10^{-5}$    & $3\times 10^{-5}$     \\
    Wolf-Rayet  & 0        & 0.62      & 0.37     & 0.01      & 0         \\ 
    \bottomrule
\end{tabular*}
\end{table*}

\subsection{Emitted Power}
\label{sec:cr_emission}

Assuming that the cosmic-ray acceleration occurs in the relativistic jet, the cosmic-ray luminosity can be estimated as a fraction of the total power of the jet. The cosmic-ray emission rate per energy unit was written as
\begin{align}
\frac {dN_j} {dEdt} = \eta f_j \frac{Q_{\text{jet}}}{E^{\text{cr}}_{\text{jet}}} \left(\frac{dN}{dE}\right)_{\text{model}},
\end{align}
where  $f_j$ is the fraction of nuclear species $j$ at the source, $Q_{\text{jet}}$ is the total jet kinetic power, and $\eta$ is an effective normalization factor relating the arriving flux fraction and the jet power. The normalization factor
\begin{align} \label{normfactor}
E^{\text{cr}}_{\text{jet}} = f_H\int_{E_0}^{E^*} dE E \left(\frac{dN}{dE}\right)_{\text{DSA}} + \nonumber\\
\sum_jf_j\int_{E^*}^\infty dE E \left(\frac{dN}{dE}\right)_{\text{model}},
\end{align}

\noindent accounts for the total energy in cosmic rays. The lower energy bound in the integral, $E_{0}$, is adopted as $E_0 = A m_p c^2$. The upper limit is approximated as infinity due to the exponential cutoff.

Based on the acceleration timescales derived by Seo et al.~\cite{Seo_2023}, we assume that for energies below $E^* = 10^{17} \text{ eV}$, the spectrum will be dominated by the standard DSA mechanism with $s=2$. Although the acceleration timescale for nongradual shear acceleration is shorter than that for DSA, the former is only operative when the scattering mean free path is large enough to cross the jet-backflow interface~\cite{Seo_2023}. Since the composition content of AGN jets remains uncertain, we assume the hadronic component is largely dominated by protons for energies below $E^*$ as a simplification~\cite{Celotti_1993}.
\subsection{Source distribution}
\label{sec:continuous_sources_model}

The most prominent individual candidates are the nearby FR-I radio galaxies Centaurus A (Cen~A, NGC 5128, 3.8 Mpc), Virgo A (Vir~A, M87, 16.4 Mpc), and Fornax A (For~A, NGC 1316, 20.8 Mpc)~\cite{Harris_Rejkuba_Harris_2010,2018kobzar,Cantiello_2013}. These objects are widely recognized in the literature as the most promising local UHECR sources, based on their favorable combination of substantial jet power and proximity to Earth~\cite{Matthews_2018, Eichmann_2018, Lang_2020, de_Oliveira_2023, Biermann_2012}. Their individual treatment is further motivated by their angular proximity to UHECR excesses reported by large-scale experiments, which has led several studies to investigate their potential contributions in detail~\cite{Matthews_2018, Mollerach_2024}. Nevertheless, these three radio galaxies are not expected to be the sole contributors to the observed UHECR flux. We therefore also include, in our calculations, the cumulative background contribution from the more numerous and distant population of radio galaxies. In the following, we refer to Cen~A, Vir~A and For~A as \textit{nearby sources} and to the others as the \textit{background contribution}.

The background contribution was estimated using a continuous distribution of sources~\cite{Eichmann_2018}. The UHECR emissivity was derived from the continuous source function $\Psi(E, z)$, expressed in terms of the radio luminosity $L$, energy $E$, and redshift $z$,
\begin{align}
    \Psi_j(E, z) \equiv \frac{dN_j}{dVdEdt} = \nonumber\\
    \int d L \;
    \frac{dN_{\text{RG}}}{dV\,d L}(L, z)
    \left[\frac{dN_j}{dEdt}(E, Q_{\text {jet}}(L))\right],
    \label{eq:general_continuous_source_function}
\end{align}
for each nuclear species $j $ with atomic number $Z$.

The first term in the equation~\ref{eq:general_continuous_source_function} represents the number of radio galaxies per luminosity per volume, and is estimated through radio luminosity functions (RLFs). A recent updated RLF for local FR-I radio galaxies (separately from FR-II) with $z<0.3$ was obtained by Jong et al. \cite{Jong_2024} as
\begin{equation}
\label{eq:jong_radio_luminosity_function}
    \frac{dN_{\text{RG}}}{dV\,d\log L} =
    \rho_0 \left[
    \left(\frac{L_\nu}{\bar L}\right)^\alpha +
    \left(\frac{L_\nu}{\bar L}\right)^\beta
    \right]^{-1},
\end{equation}
with $\rho_0 = 10^{-5.5} \, \text{Mpc}^{-3} \, d \log L^{-1}$, $\alpha = 0.59$, $\beta = 2.0$, and  $\bar L = 10^{25.8} \, \text{W Hz}^{-1}$. For higher redshifts, we use the straightforward evolution $\Psi_j(E,z) = \Psi_j(E, z = 0)\,\varepsilon(z)$ with $\varepsilon(z) \propto (1 + z)^m$.  Following Refs.~\cite{Batista_2019, Gelmini_2012, 2012_Ballo}  we adopt a piecewise evolution appropriate for medium-to-high luminosity AGNs, setting $m = 5$ for $z < 1.7$, $m = 0$ for $1.7 < z < 2.7$, and $m = 2.7 - z$ for $z > 2.7$.

The second term in equation~\eqref{eq:general_continuous_source_function} is the energy spectrum injected by the source, as described in section~\ref{sec:acceleration_models}. In the absence of specific parameters for each source, we approximate $\Gamma_{\rm eff} \approx 2$~\cite{Ye_2023,Kleijn_2002} and use Appendix \ref{appendix:break_energy} to predict $E_{\rm break}$ from $Q_{\rm jet}$. Using the empirical relation between jet power and the $151$~MHz radio luminosity~\cite{Willott_1999}, eq.~\eqref{eq:general_continuous_source_function} is written in terms of $Q_{\rm jet}$ as
\begin{equation}
\Psi_j(E,z) = \frac{7}{6}
\eta f_j\rho_0 \varepsilon(z) I(E),
\end{equation}
with 
{
\small
\begin{align} 
I(E) =  
\int_{Q^{\min}_{\text{jet}}}^{Q^{\max}_{\text{jet}}} \frac{dQ_{\text{jet}}} {\left(\dfrac{Q_{\text{jet}}}{f^{3/2}\bar{Q}}\right)^{7\alpha/6} + \left(\dfrac{Q_{\text{jet}}}{f^{3/2}\bar{Q}}\right)^{7\beta/6}} \nonumber\\
\left[\frac{1}{E^{\text{cr}}_{\text{jet}}} \left( \frac{dN}{dE} \right)_{\text{model}} \right]\Biggr|_{(E, E_{\text{break}}(Q_{\text{jet}}))},
\end{align}
}

\noindent where $(dN/dE)_{\text{model}}$ and $E^{\text{cr}}_{\text{jet}}$ are as defined in sections~\ref{sec:acceleration_models} and \ref{sec:cr_emission}, $f_j$ is the fraction of nucleus $j$ in the composition, and $\bar Q \equiv Q_{\text{jet}}(\bar L)/f^{3/2}$. The factor $1 \leq f \leq 20$ accounts for uncertainties in the conversion from radio luminosity to jet power. We adopt $f \sim 10$, which provides a reasonable order-of-magnitude estimate for FR-I radio galaxies~\cite{Cao_2004}.

The integration limits correspond to the least and most powerful radio galaxies considered. We set $(Q_{\text{min}}, Q_{\text{max}}) = (10^{43}, 2\cdot 10^{46}) \text{erg s}^{-1}$ to be consistent with limits used by Jong \textit{et al} \cite{Jong_2024}. Based on our numerical tests and the results from Eichmann et al \cite{Eichmann_2018}, the observed UHECR flux depends only weakly on $Q_{\text{min}}$. 

\section{Simulation setup and data analysis method}
\label{sec:methods}
\subsection{Extragalactic propagation}
\label{sec:propagation}

We simulate the extragalactic propagation of astroparticles using the \crpropa\ Monte Carlo framework~\cite{Batista_2022}, assuming a one-dimensional geometry in the absence of magnetic fields. We set the sources to emit a power-law spectrum with spectral index $-1$ over the range $\left[E_{\text{min}}, E_{\text{max}}\right] = \left[10^{18}, 10^{21}\right] \text{eV}$, providing a uniform number of events per energy decade. The injected composition included equal fractions of $^1\text{H}$, $^4\text{He}$, $^{14}\text{N}$, $^{28}\text{Si}$, and $^{56}\text{Fe}$ primaries. The energy spectra explained in section~\ref{sec:acceleration_models} are introduced in the data analysis step via the weighting procedure explained in Appendix~\ref{appendix:normalization}.

The \crpropa \ framework accounts for photo-pion production, photo-pair production, and photo-disintegration during UHECR propagation, utilizing the cosmic microwave background (CMB) and the Gilmore~et~al. extragalactic background light (EBL) model~\cite{Gilmore_12} as photon targets. The simulations also include nuclear decay processes and adiabatic energy losses driven by cosmic expansion. The cosmological parameters considered are $H_0 = 67.3\ \text{km s}^{-1}\text{Mpc}^{-1}$, $\Omega_m = 0.315$, and $\Omega_\Lambda = 1 - \Omega_m$~\cite{Olive_2014}. Simulation tracking cutoffs are set at $0.9\ \text{EeV}$ for nuclei and $0.1\ \text{PeV}$ for neutrinos.

We model Cen~A, Vir~A, and For~A as individual point-source emitters, while the background contribution is simulated as a uniform distribution of sources within a 1D shell-like geometry bounded by $D_{\text{min}}$ and $D_{\text{max}}$. The comoving distances for these shells are logarithmically spaced from $25~\text{Mpc}$ out to $4~\text{Gpc}$. For each point source and background shell, we simulate $N = 10^6$ primary events. Further details on the normalization procedure are provided in Appendix~\ref{appendix:normalization}.

\subsection{Procedure to fit the measured energy spectrum}
\label{sec:fitting}

Our model incorporates four separate source contributions: i) Cen~A, ii) Vir~A, iii) For~A and iv) background. We fit this model to the energy spectrum measured by the Pierre Auger Observatory~\cite{Auger_2020} by optimizing the flux fractions of each component to best describe the data. Consequently, the free parameters of the fit are the relative normalization fractions of the three discrete sources and the background. Mass composition observables are excluded from the fitting procedure, serving instead as independent\textit{ a posteriori} consistency checks for our best-fit scenarios. Incorporating composition observables directly into the fit is left for future work.

We implement our fitting procedure using the CERN \texttt{ROOT} framework~\cite{2025_ROOT} and the \texttt{Minuit2} minimizer by minimizing the deviance $D = - 2 \ln(L/L_{\rm sat})$. Here, $L$ is the model likelihood and $L_{\rm sat}$ is the likelihood of a model that perfectly describes the data~\cite{Auger_2023, 1984_Baker}. We define the total likelihood function as a product of Gaussian distributions for the energy bins where the Pierre Auger Observatory detected events, and a negative Poisson log-likelihood for bins consisted by upper limits. The total deviance is  given by $D=\sum_i d_i$, where for each $i$-th energy bin,
\begin{equation}\label{eq:deviance_function}
        d_i = \begin{cases}
                    (J_i^{\rm obs}-J_i^{\rm mod})^2/\sigma_i^2, & \text{if }i\text{ is a measurement};  \\
                    2 J_i^{\rm mod}/J_i^{\rm un}       & \text{if }i\text{ is an upper limit.}    
              \end{cases}
\end{equation}
Here $J_i^{\rm obs}$ is the observed flux and $J_i^{\rm mod}$ is the sum of the flux contribution from Cen~A, Vir~A, For~A and the background. The quantity $\sigma_i^2$ denotes the statistical variance of the data; in the case of asymmetric uncertainties, we adopt the larger value. 

For bins with zero observed events (reported as $90\%$ confidence level upper limits~\cite{Auger_2020}), the Poisson deviance is reduced to $d_i=2\mu_i$. The expected counts are written as $\mu_i = J_i^{\rm mod}/J_i^{\rm un}$, where $J_i^{\rm un}$ is the flux corresponding to one event in that bin. For an energy bin of width $\Delta E_i$ and exposure $\mathcal{E}$, this quantity is
\begin{equation}
    J_i^{\rm un} = \frac{c_i}{\Delta E_i \mathcal{E}},
\end{equation}
where $c_i$ is a correction factor \cite{Aab2020-lb}. The final goodness-of-fit statistic is the deviance per degree of freedom, $D/\text{ndf}$, where the number of degrees of freedom $(\text{ndf})$ is given by the number of fitted data points minus the number of fit parameters.


\section{Results}
\label{sec:results}
\subsection{UHECR fitted energy flux and resulting composition}
\label{sec:UHECR_energy_fitting}

The framework developed here enables us to investigate several key questions, including the relative contribution of nearby sources with respect to the background, the influence of primary composition, and the shape and characteristic parameters of the energy spectrum. To address these questions, we define different combinations of model parameters, referred to as Scenarios, as summarized in Table~\ref{tab:case-definitions}. 
 
For each Scenario, the corresponding prediction of the mean logarithmic mass number $\langle \ln A \rangle$ obtained at Earth is shown in comparison with measurements from the Pierre Auger Observatory \cite{PhysRevD.111.022003}. The reported values of $\langle \ln A \rangle$ correspond to the Auger data points shown in the figure and are derived from measurements of the depth of maximum of air-shower profiles, $X_{\rm max}$ ~\cite{PhysRevD.111.022003,Aab_2021}. Their results depend on the choice of hadronic interaction model used in the analysis: QGSJet~\cite{Ostapchenko_2011}, EPOS~\cite{Pierog_2015} or Sibyll~\cite{Riehn_2020}. 

Our model predictions for $\langle \ln A \rangle$ are obtained by computing the mean logarithmic mass number from the resulting nuclear mass fractions arriving at Earth. Because the model is fitted exclusively to the energy spectrum measured by the Pierre Auger Observatory, $\langle \ln A \rangle$ is not included as a fit observable; instead, it is computed \textit{a posteriori } using the best-fit source parameters that determine the relative abundances of the arriving nuclei.

\begin{table*}[t]
\caption{Scenarios tested in this work.}
\label{tab:case-definitions}

\begin{tabular*}{\textwidth}{@{\extracolsep\fill}cccccc}
\toprule
    Scenario & Energy spectrum & s &  With background sources? & Fit range ($\rm EeV$) \\ \midrule
    1.a    & DPLEC  & -   &  No   &    $20-200$      \\
    1.b    & DPLEC  & -   &  Yes  &    $20-200$      \\
    2.a    & PLEC   & 2.0 &  No   &    $5-200$       \\
    2.b    & PLEC   & 2.0 &  Yes  &    $5-200$       \\
    3      & PLEC   & 2.3 &  Yes  &    $5-200$       \\
    \bottomrule
    \end{tabular*}

\end{table*}

\begin{figure*}[t]
        \begin{minipage}[b]{0.49\linewidth}
            \centering
            \includegraphics[width=\textwidth]{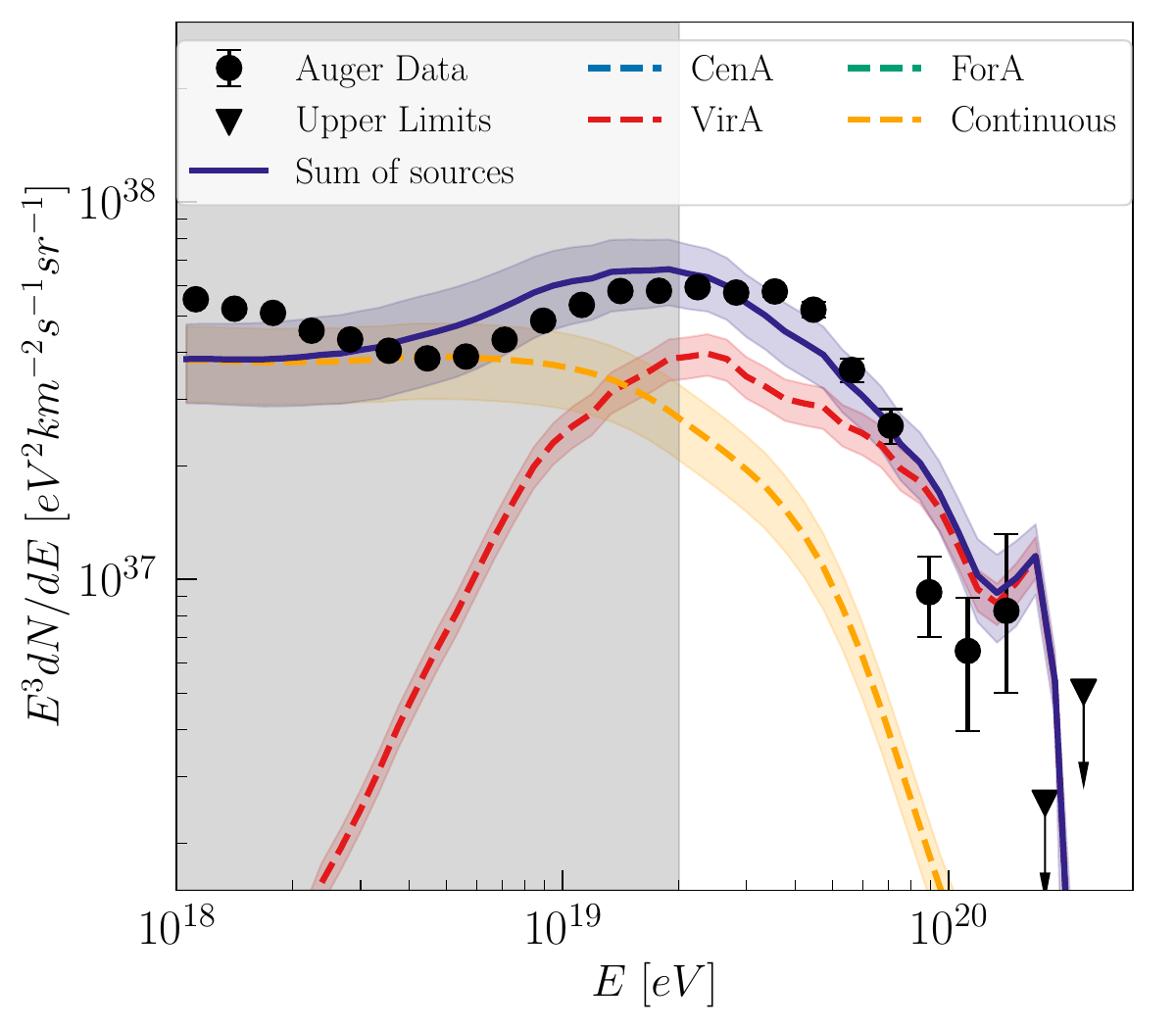}
        \end{minipage}
        \hspace{0.5cm}
        \begin{minipage}[b]{0.49\linewidth}
            \centering
            \includegraphics[width=\textwidth]{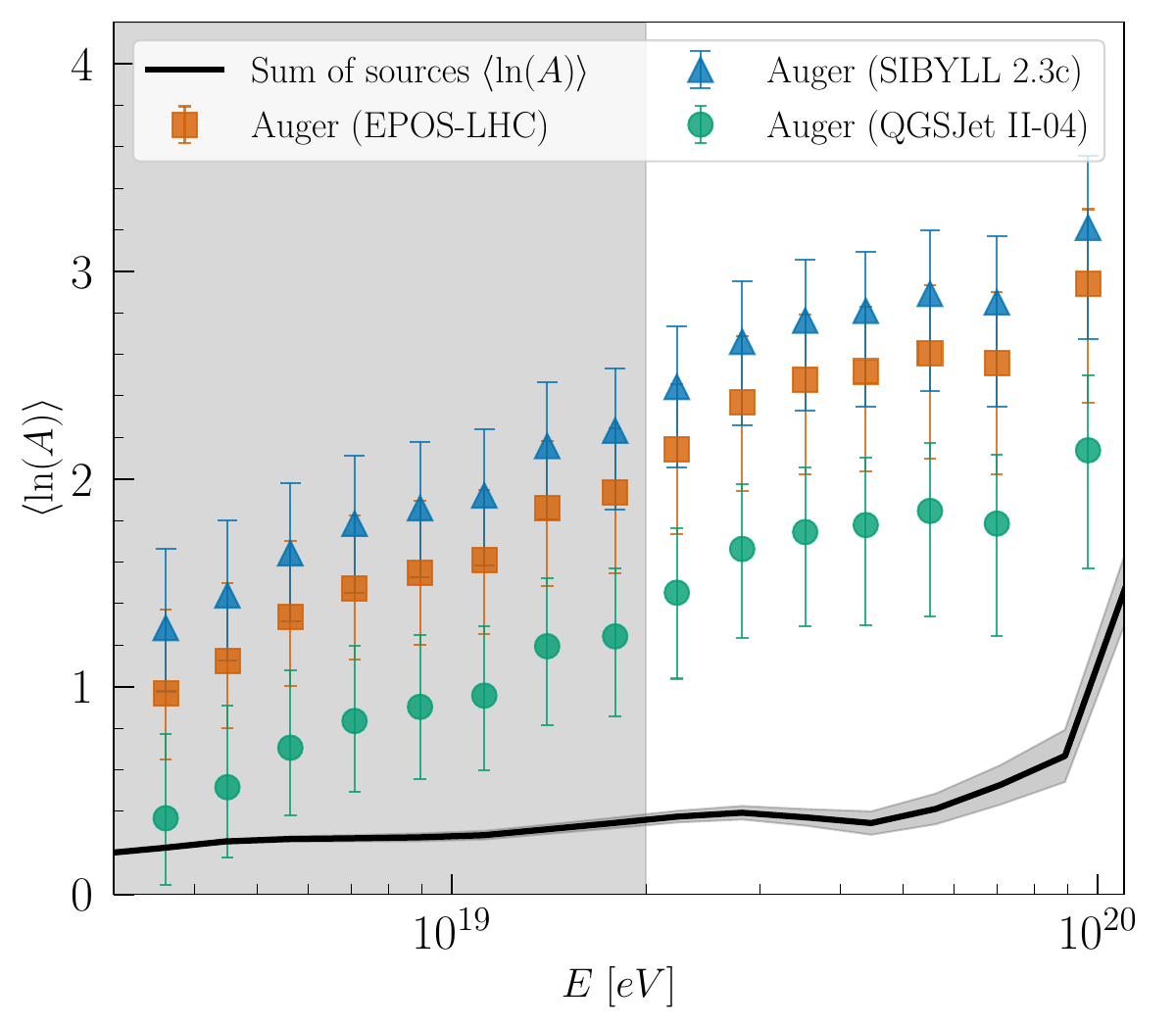}
        \end{minipage} 
        \caption{Results for Scenario $1.\rm b$ using \textit{Solar} composition. \texttt{Left}: Energy spectrum of UHECRs compared with the Pierre Auger Observatory data~\cite{Auger_2020}. Contributions from individual nearby sources, a continuous background source distribution (dashed yellow line) beyond the nearby radio galaxies and their sum are given respectively by dashed and solid lines with shaded bands (statistical uncertainties of the best-fit parameters). \texttt{Right}: Corresponding mean logarithmic mass number $\langle \ln A \rangle$ predicted by the summed source contribution (black solid line), compared with the Pierre Auger Observatory data~\cite{Auger_2023} interpreted with different hadronic interaction models: EPOS-LHC, SIBYLL 2.3c and QGSJet II-04. Shaded vertical regions in both panels indicate energy ranges excluded from the fit.}
        \label{fig:friall_solar}
\end{figure*}

For all scenarios considered, the model using \textit{Solar} composition was unable to reproduce the UHECR energy spectrum measured by the Pierre Auger Collaboration. Because the fit relies exclusively on the energy spectrum, the poor goodness-of-fit obtained for this configuration ($D/\mathrm{ndf}\sim10$--50) is sufficient to disfavor the Solar assumption on spectral grounds alone. Furthermore, the corresponding predicted evolution of the mean logarithmic mass is very light and nearly flat, which is found \emph{a posteriori} to be highly inconsistent with composition measurements (illustrated for Scenario~1.b in Figure~\ref{fig:friall_solar}). Consequently, configurations adopting a Solar composition are excluded from further analysis, and we adopt the Wolf-Rayet composition for the remaining results. 

The best-fit parameter sets, corresponding deviance values and p-values for each Scenario are presented in Table~\ref{tab:resultstable}, alongside references to their respective figures. Physically, the parameter $\eta$ represents the fraction of jet power converted into non-thermal cosmic rays. In practice, however, it must be treated as an effective normalization factor. Jet powers are determined indirectly using proxies that can yield estimates varying by up to two orders of magnitude~\cite{Foschini_2024}. Furthermore, because AGNs exhibit intrinsic variability, the UHECR flux currently arriving at Earth likely reflects past historical states of the source~\cite{Matthews_2018, deOliveira_2025_history}. Finally, deflections by coherent EGMFs~\cite{de_Oliveira_2022, de_Oliveira_2023} and the GMF~\cite{Bister_2024} can supress or amplify the UHECR flux from any individual source. Therefore, the fitted values of $\eta$ should be interpreted as effective normalization parameters that relate the observed UHECR flux to the estimated jet power, rather than as direct measurements of the physical acceleration efficiency. They incorporate uncertainties associated with the jet-power estimates, source variability, magnetic-field effects, and other simplifying assumptions of the model. Consequently, the relative values of $\eta$ between different scenarios or sources are more robust than their absolute numerical values.

\begin{table*}[t]
\caption{Best-fit parameters for each Scenario.}
\label{tab:resultstable}

\begin{tabular*}{\textwidth}{@{\extracolsep\fill}cccccccc}
\toprule
    Scenario & Figure                    & $\eta_{\rm Cen~A} (10^{-3})$ & $\eta_{\rm Vir~A}  (10^{-2})$        & $\eta_{\rm For~A}$ & $\eta_{\rm Cont} (10^{-3})$        & $\text{D/ndf}$ & $\text{p-value}$\\ \midrule
    $1.\rm a$ & \ref{fig:frinearby}      & $3.7\pm0.9$ & $0\pm20$      & $5.4\pm0.3$   & -            & $19.55/8$  & $0.012$ \\
    $1.\rm b$ & \ref{fig:friall}         & $6\pm1$     & $0\pm2$       & $3\pm1$       & $8\pm3$      & $6.16/7$   & $0.52$ \\
    $2.\rm a$ & \ref{fig:fermi2.0nearby} & $1.8\pm0.1$ & $0.00\pm0.04$ & $1.50\pm0.04$ & -            & $84.52/14$ & $10^{-12}$\\
    $2.\rm $b & \ref{fig:fermi2.0all}    & $2.5\pm0.2$ & $0\pm7$       & $1.17\pm0.07$ &$0.25\pm0.04$ & $13.08/13$ & $0.44$ \\
    $3$ & \ref{fig:fermi2.3all}          & $2.6\pm0.7$ & $1.3\pm0.4$   & $0.46\pm0.05$ &$0.00\pm0.01$ & $25.47/13$ & $0.02$ \\ \bottomrule
\end{tabular*}

\end{table*}

\begin{figure*}[t]
        \begin{minipage}[b]{0.49\linewidth}
            \centering
            \includegraphics[width=\textwidth]{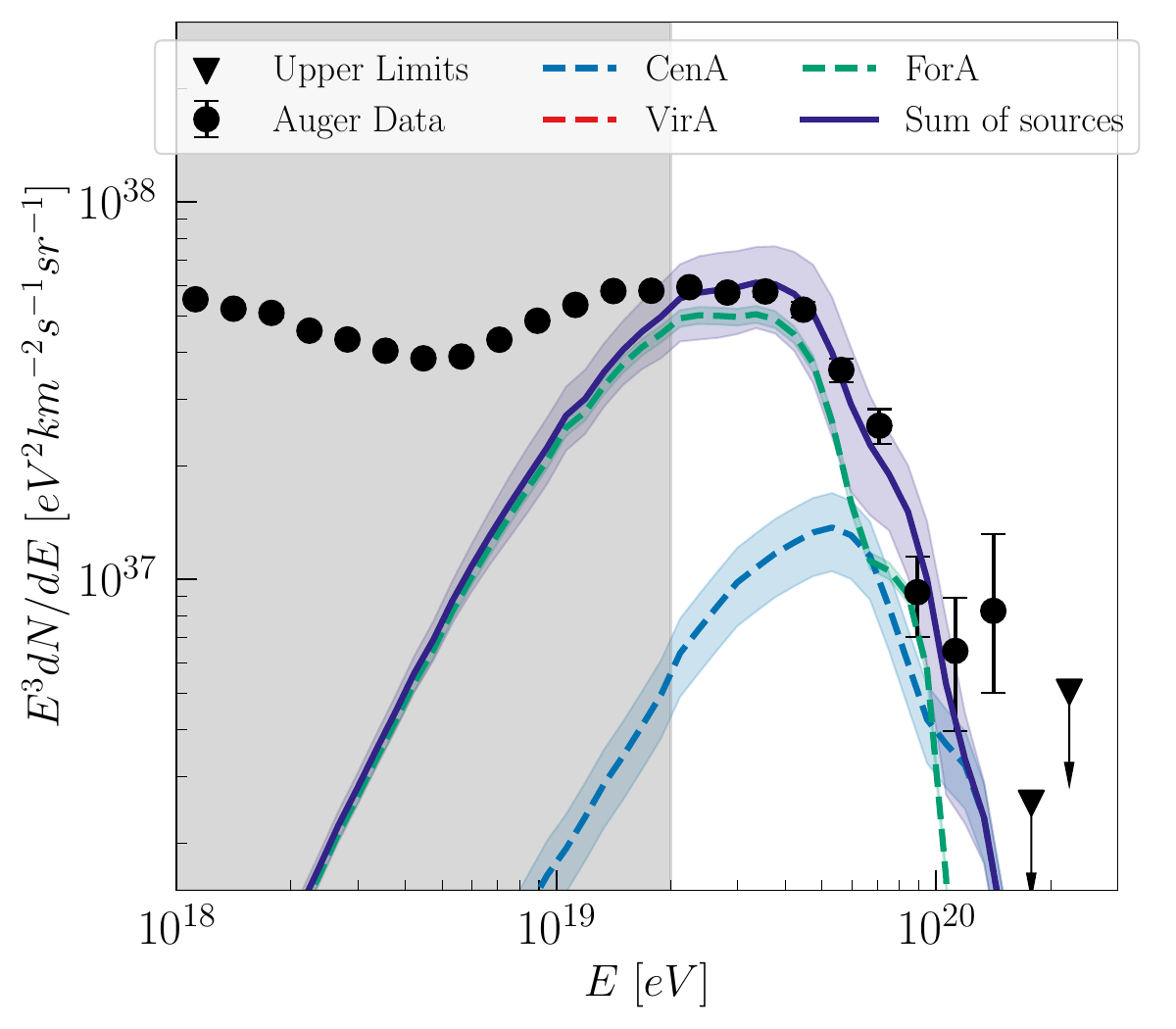}
        \end{minipage}
        \hspace{0.5cm}
        \begin{minipage}[b]{0.49\linewidth}
            \centering
            \includegraphics[width=\textwidth]{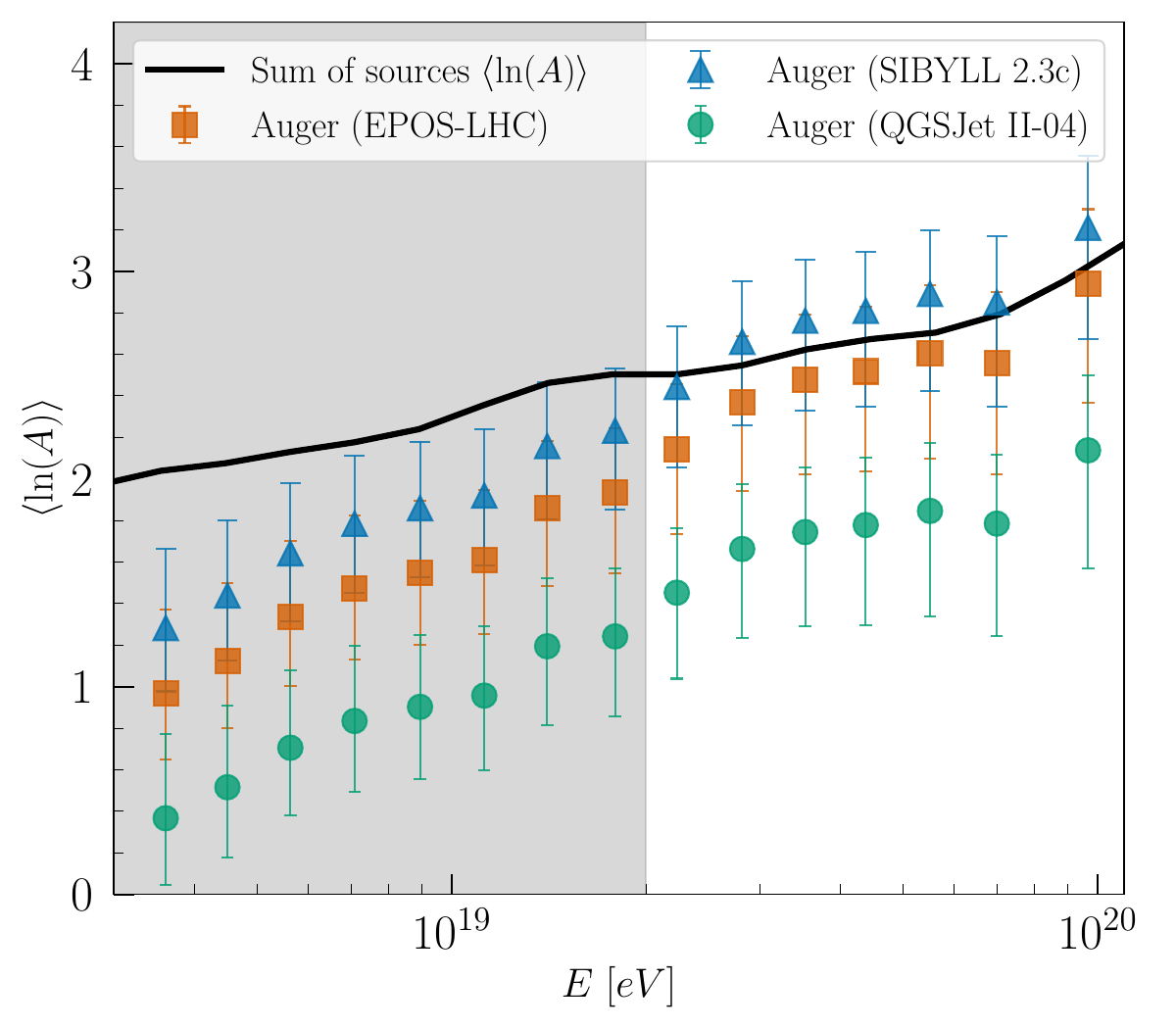}
        \end{minipage}
        \caption{Resulting UHECR spectrum (\texttt{left}) and corresponding $\langle \ln A \rangle$ (\texttt{right}) for Scenario $1.\rm a$. The elements are the same as Figure \ref{fig:friall_solar}, but without the continuous source distribution beyond the nearby radio galaxies.}
        \label{fig:frinearby}
\end{figure*}

\begin{figure*}[htbp]
        \begin{minipage}[b]{0.49\linewidth}
            \centering
            \includegraphics[width=\textwidth]{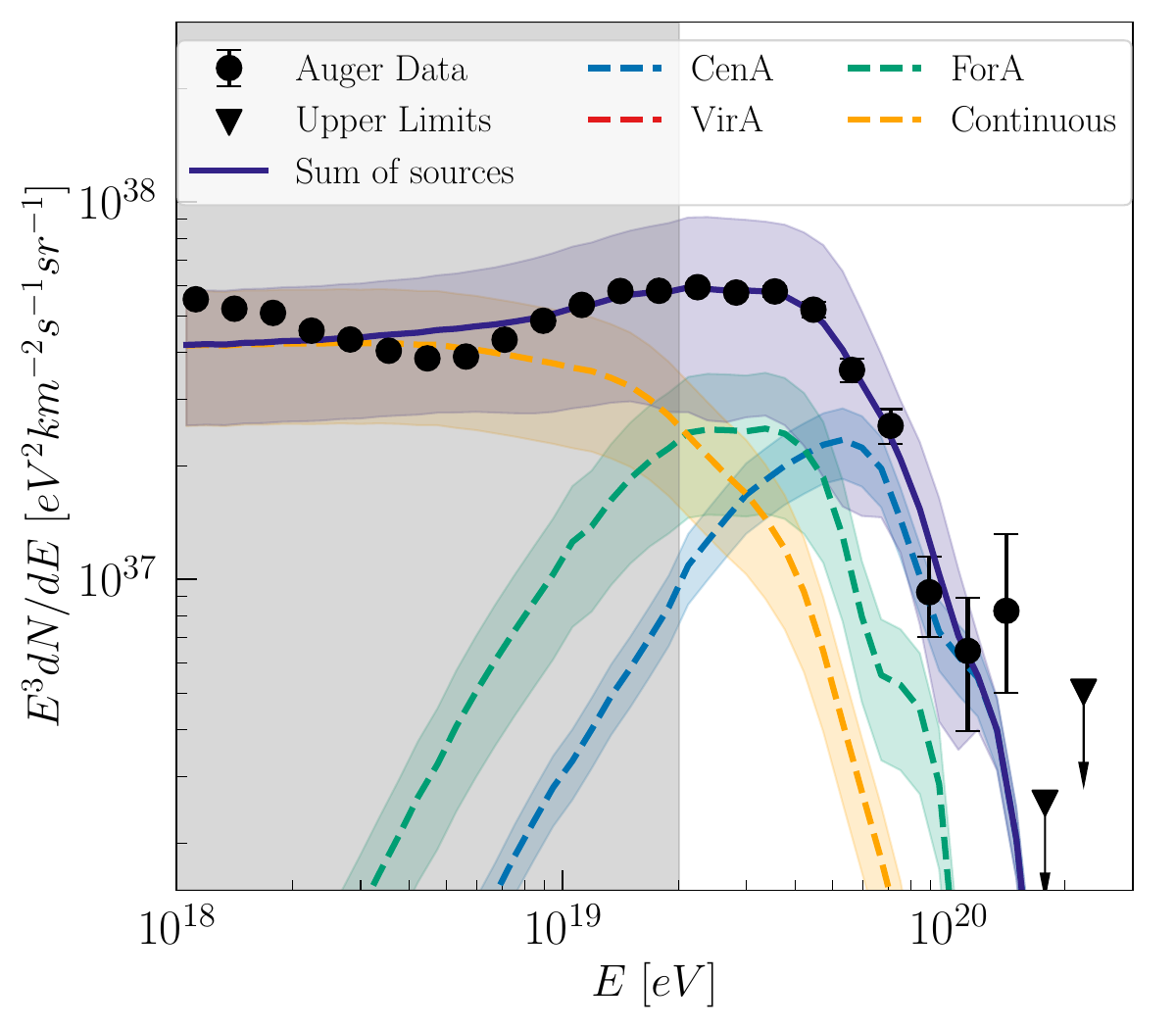}
        \end{minipage}
        \hspace{0.5cm}
        \begin{minipage}[b]{0.49\linewidth}
            \centering
            \includegraphics[width=\textwidth]{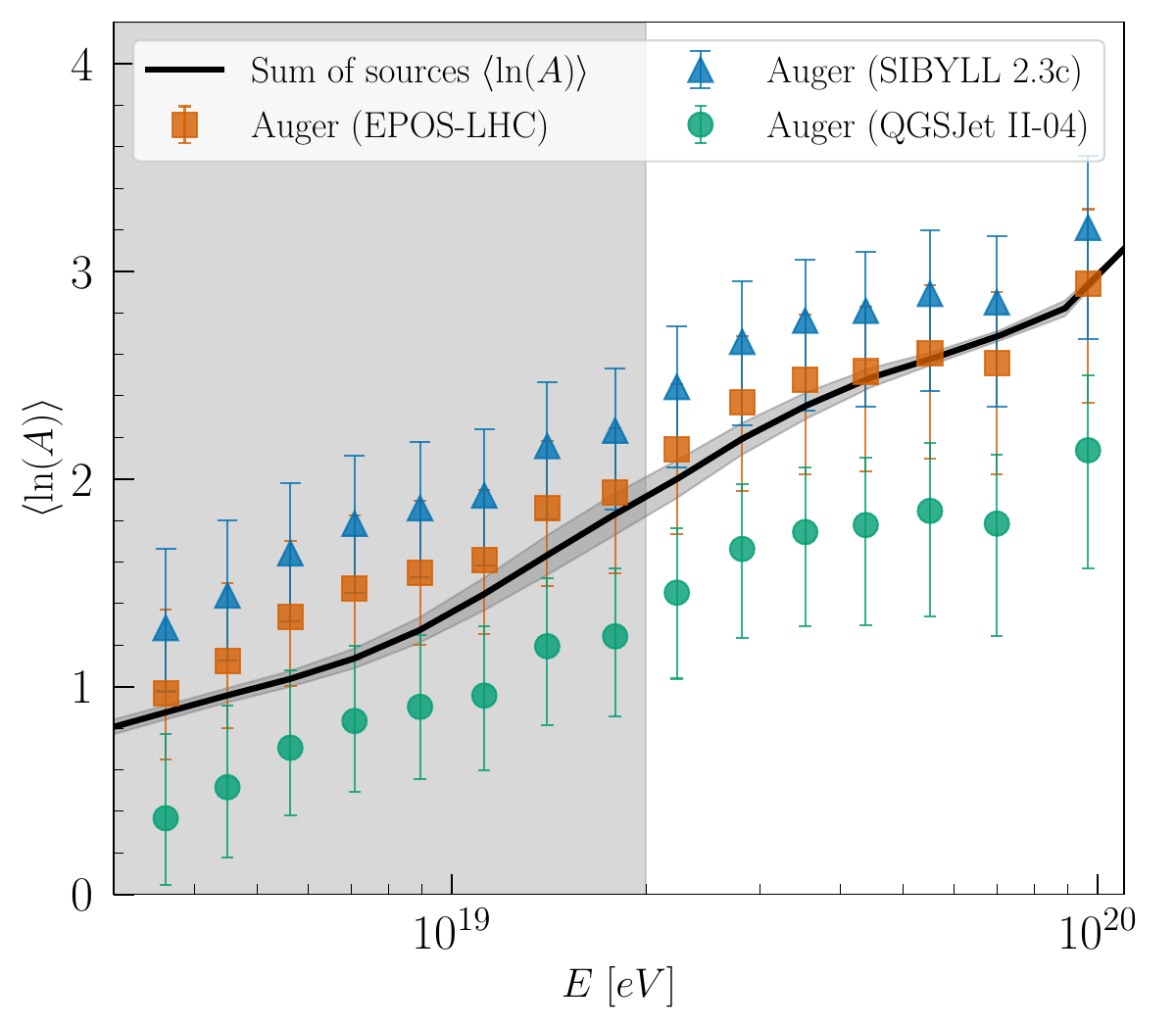}
        \end{minipage} 
        \caption{Resulting UHECR spectrum (\texttt{left}) and corresponding $\langle \ln A \rangle$ (\texttt{right}) for Scenario $1.\rm b$. The elements are the same as Figure \ref{fig:friall_solar}.}
        \label{fig:friall}
\end{figure*}

\begin{figure*}[t]
        \begin{minipage}[b]{0.49\linewidth}
            \centering
            \includegraphics[width=\textwidth]{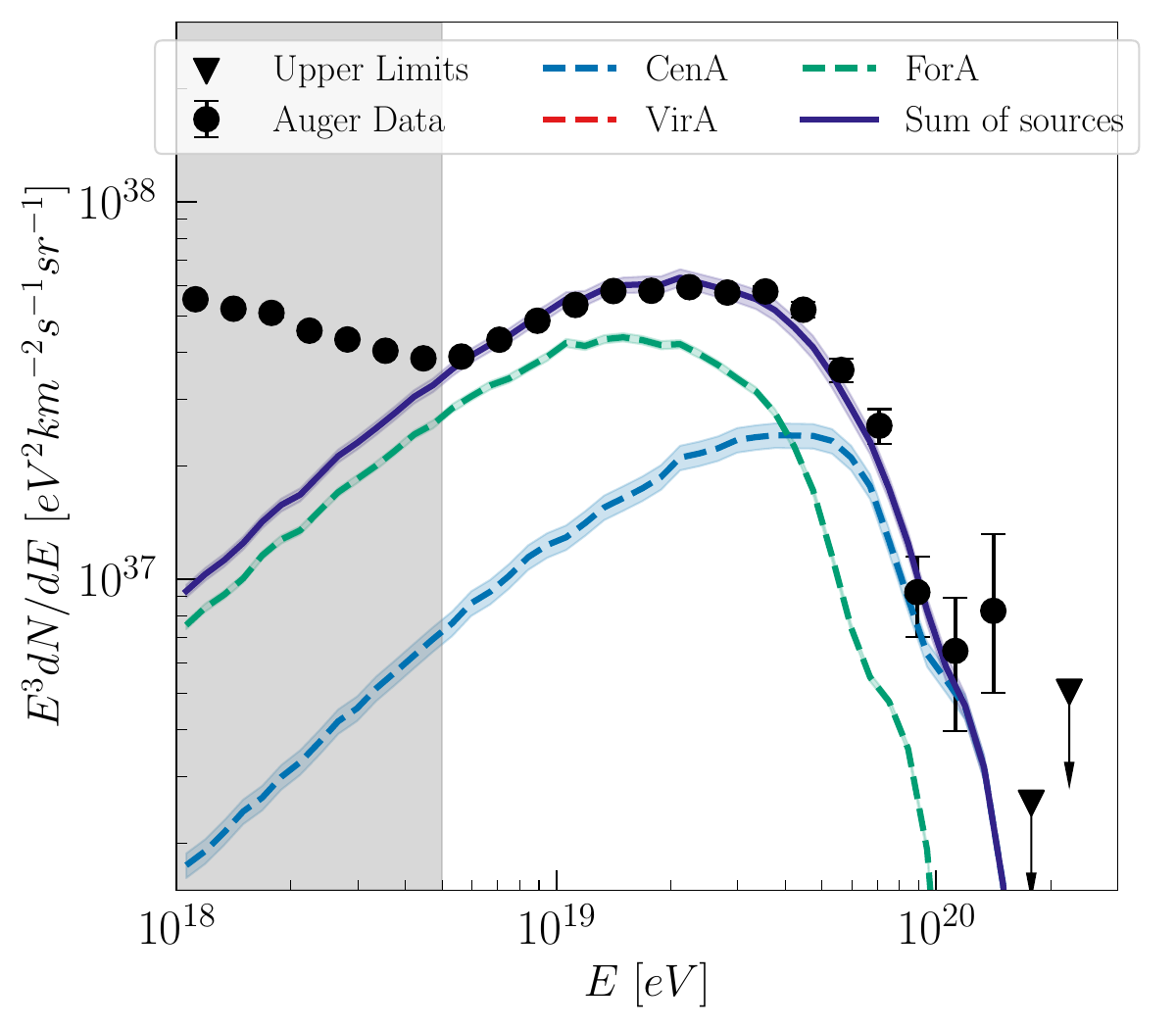}       
        \end{minipage}
        \hspace{0.5cm}
        \begin{minipage}[b]{0.49\linewidth}
            \centering
            \includegraphics[width=\textwidth]{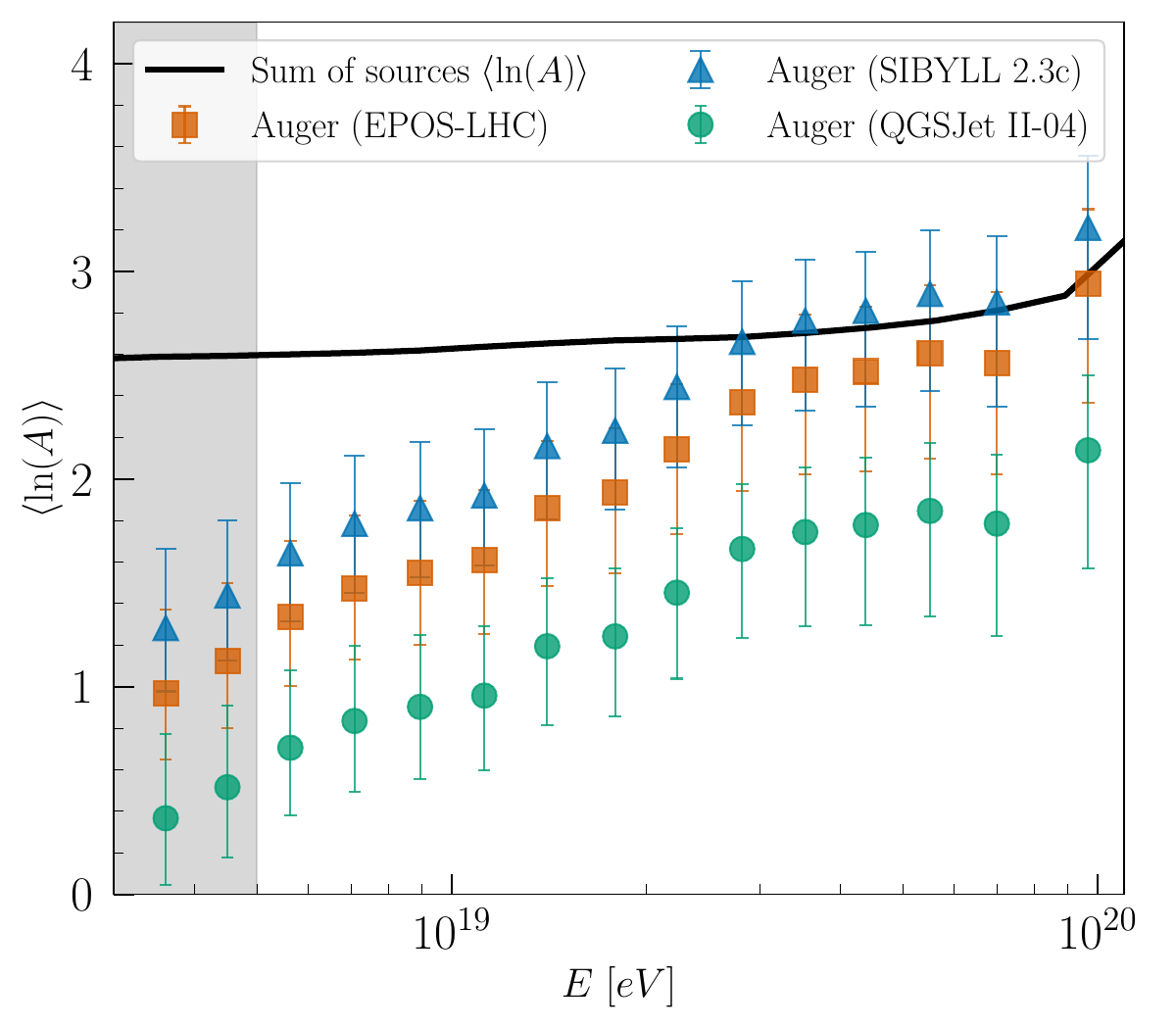}
        \end{minipage}
        \caption{Resulting UHECR spectrum (\texttt{left}) and $\langle \ln A \rangle$ (\texttt{right}) for Scenario~$2.\rm a$. The elements are the same as Figure \ref{fig:frinearby}.}
         \label{fig:fermi2.0nearby}
\end{figure*}

\begin{figure*}[t]
        \begin{minipage}[b]{0.49\linewidth}
            \centering
            \includegraphics[width=\textwidth]{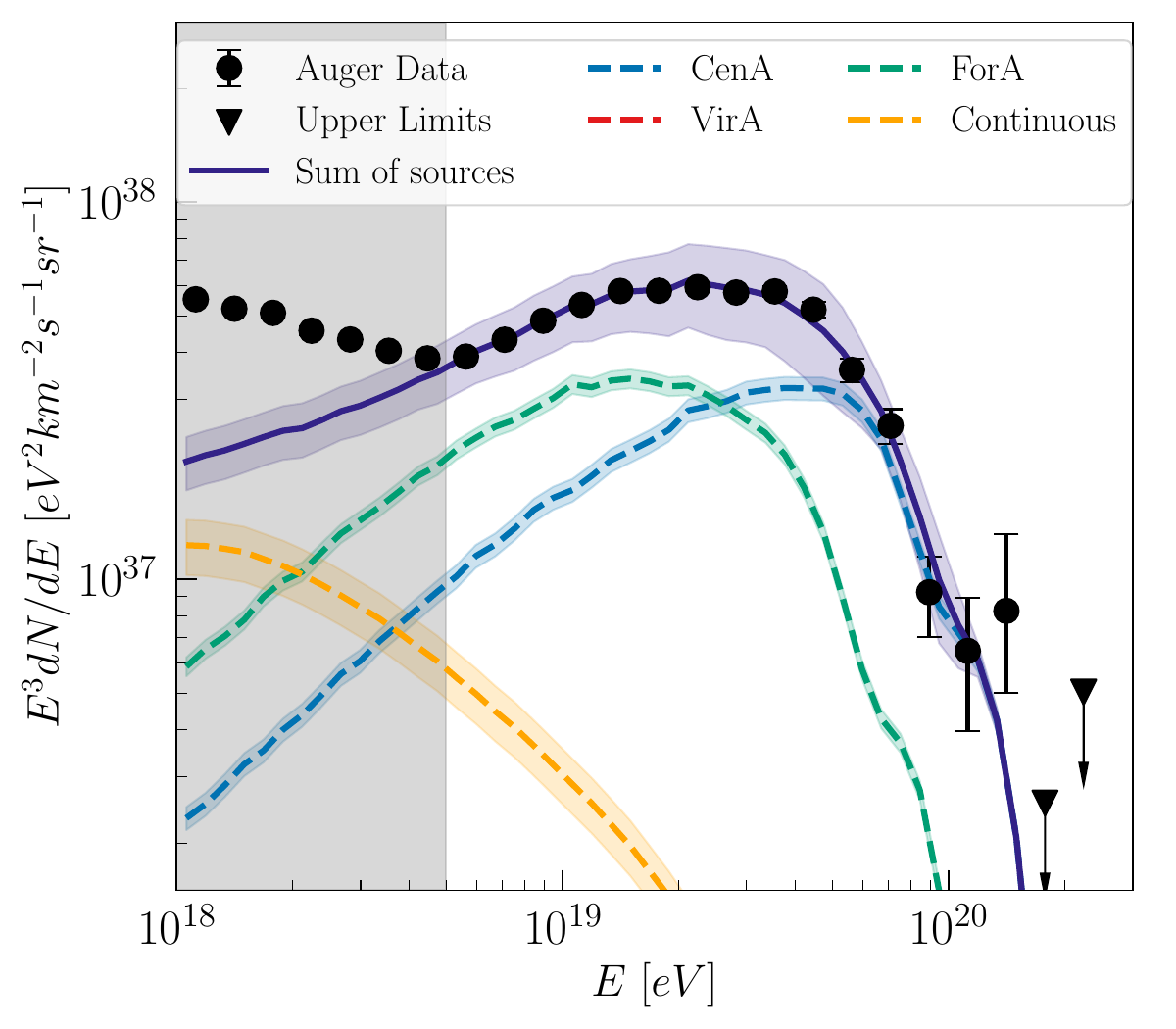}    
        \end{minipage}
        \hspace{0.5cm}
        \begin{minipage}[b]{0.49\linewidth}
            \centering
            \includegraphics[width=\textwidth]{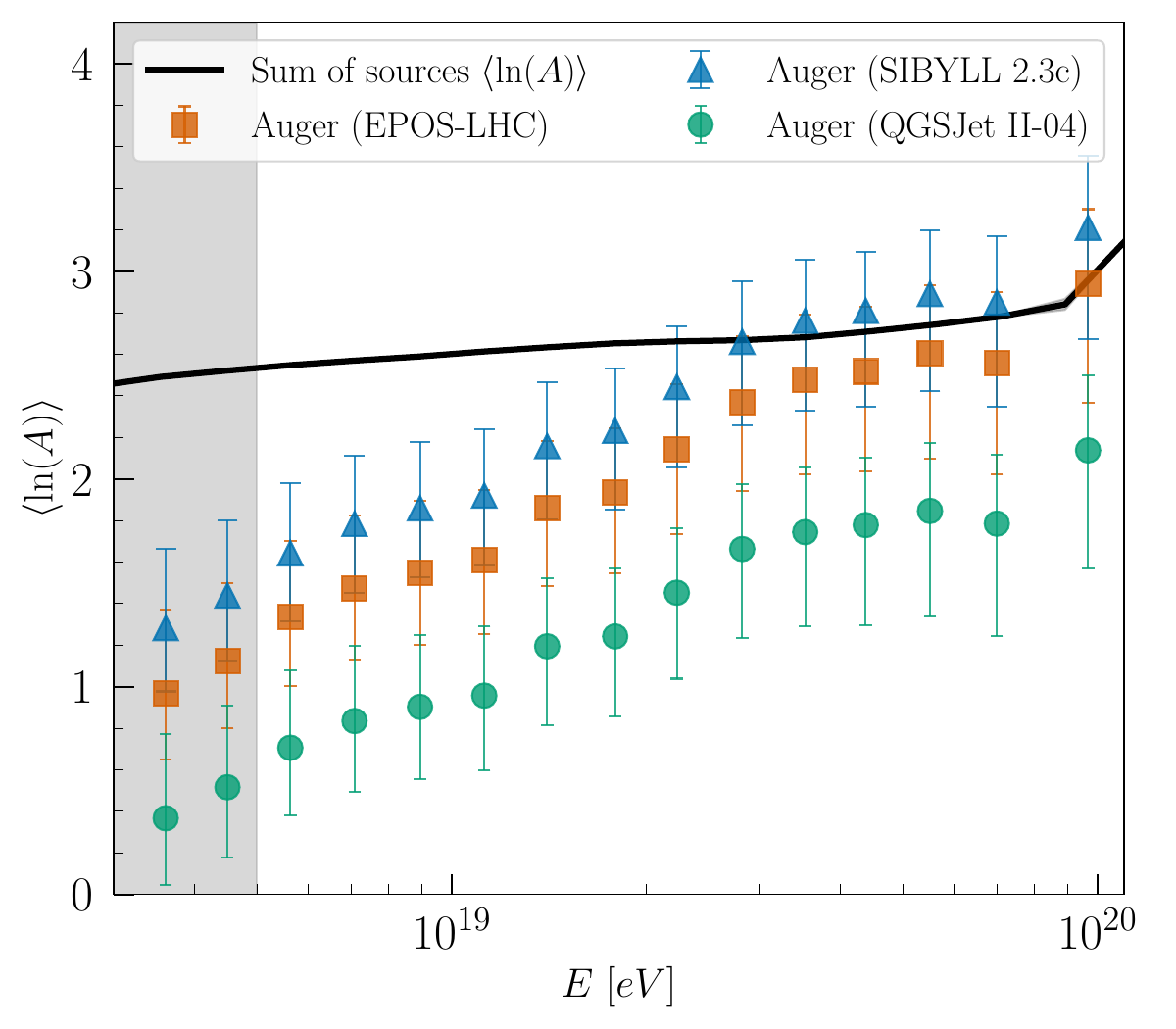}
        \end{minipage}
        \caption{Resulting UHECR spectrum (\texttt{left}) and $\langle \ln A \rangle$ (\texttt{right}) for Scenario~$2.\rm b$. The elements are the same as Figure \ref{fig:friall_solar}.}
        \label{fig:fermi2.0all}
\end{figure*}

\begin{figure*}[t]
        \begin{minipage}[b]{0.49\linewidth}
            \centering
            \includegraphics[width=\textwidth]{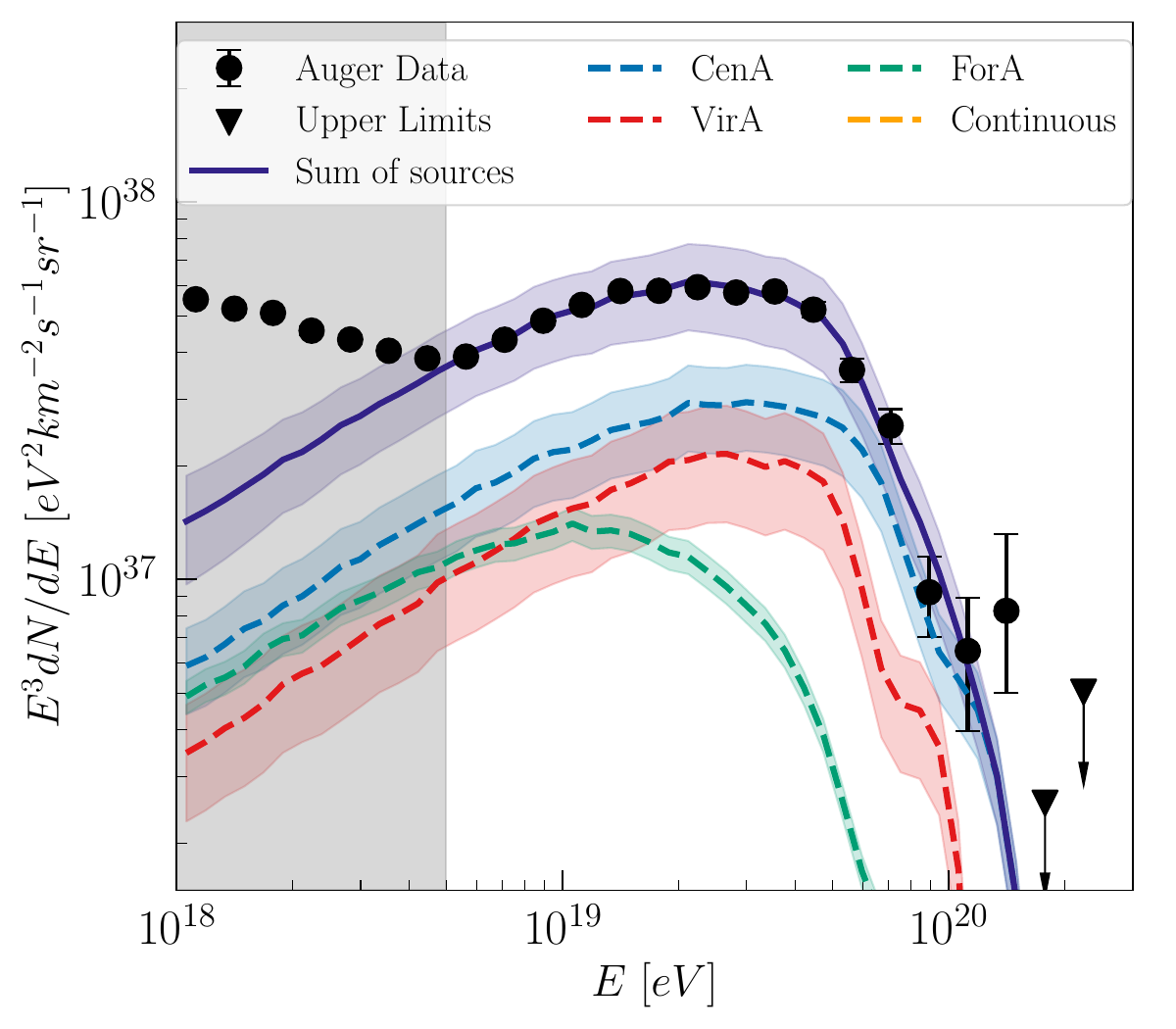}        
        \end{minipage}
        \hspace{0.5cm}
        \begin{minipage}[b]{0.49\linewidth}
            \centering
            \includegraphics[width=\textwidth]{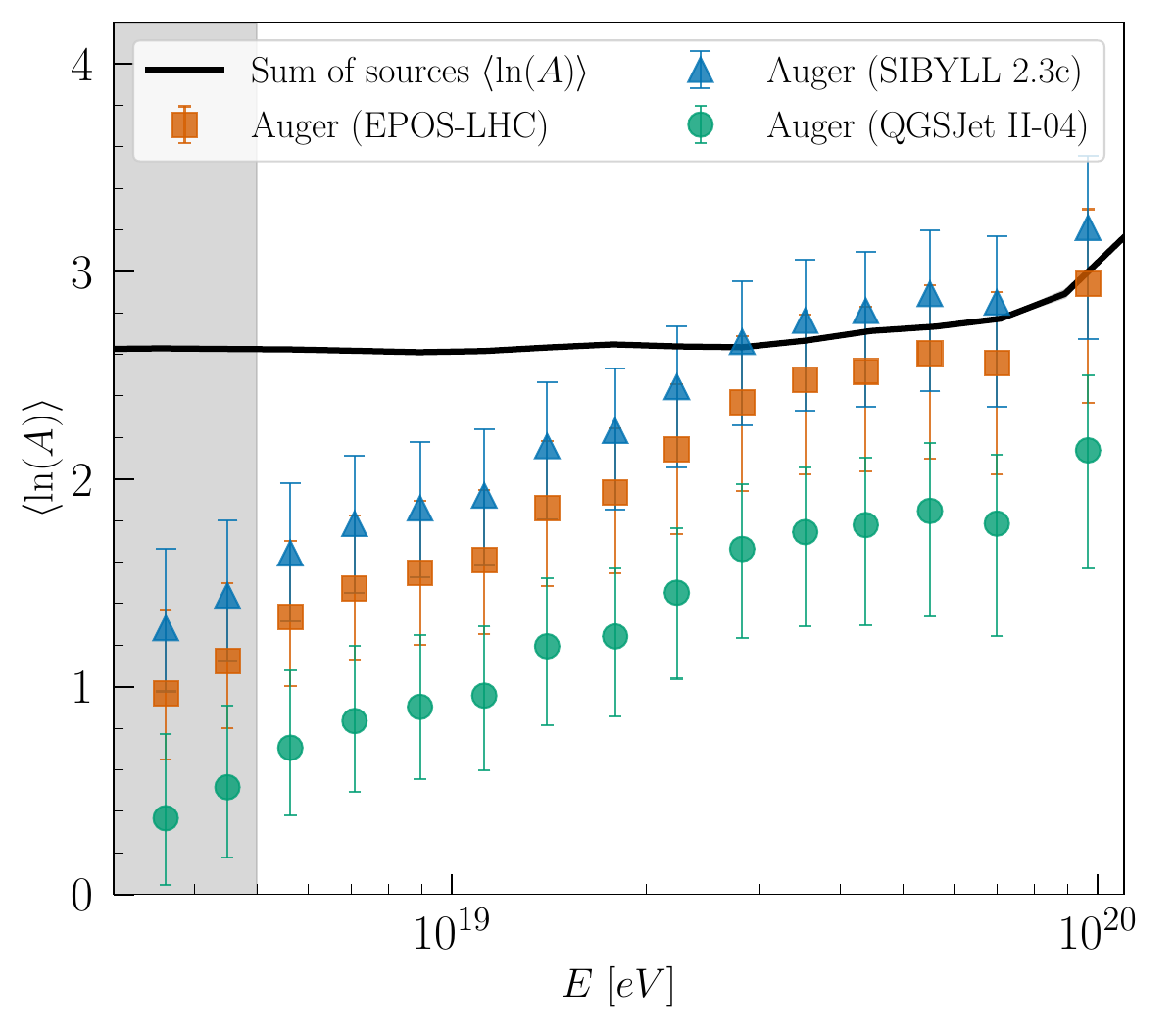}           
        \end{minipage}     
        \caption{Resulting UHECR spectrum (\texttt{left}) and $\langle \ln A \rangle$(\texttt{right}) for Scenario~$3$. The elements are the same as Figure \ref{fig:friall_solar}.}
           \label{fig:fermi2.3all}
\end{figure*}

Figures~\ref{fig:frinearby} to \ref{fig:fermi2.3all} present the energy spectrum fitted to the Pierre Auger Observatory data \cite{Auger_2020} and the respective $\langle \ln A \rangle$ for the studied Scenarios~\ref{tab:case-definitions} considering the\textit{Wolf-Rayet} composition. The shaded region indicates the energy range excluded from the fit and it should not be compared to the data. 

Figure~\ref{fig:frinearby} shows the results for Scenario~1.a. For energies above 20~EeV, Scenario~1.a describes the data reasonably: $D/\mathrm{ndf} = 19.55/8=2.44$. For~A and Cen~A provide the dominant contributions to the flux, while the best-fit contribution from Vir~A is consistent with zero within the statistical uncertainties. Although the fit is based solely on the energy spectrum, the corresponding predicted evolution of $\langle \ln A \rangle$ is found \emph{a posteriori} to be consistent with the composition measurements when the EPOS-LHC or SIBYLL~2.3c hadronic interaction models are adopted.

Figure~\ref{fig:friall} shows the results for Scenario~1.b. This Scenario was introduced to investigate the influence of the background contribution in Scenario~1.a. Including the background population improves the description of the measured energy spectrum: $D/\rm ndf = 6.16/7=0.87$. The inclusion of more distant sources causes minor changes in parameters of the nearby sources to the measured flux. For~A and Cen~A now contribute comparably to the observed flux, while the best-fit contribution from Vir~A remains consistent with zero within the statistical uncertainties. Although the fit is performed using only the energy spectrum, the inclusion of distant sources improves the \emph{a posteriori} agreement between the predicted evolution of $\langle \ln A \rangle$ and the Auger composition measurements down to $\sim 3$~EeV, compared with Scenario~1.a.

Although Scenario~1.b provides a good description of the spectrum above 20 EeV, the extrapolation toward lower energies overshoots the flux near the ankle region, indicating that this scenario cannot represent a complete description of the full UHECR spectrum.

Figures \ref{fig:fermi2.0nearby} and \ref{fig:fermi2.0all} present the results for Scenarios 2.a and 2.b, respectively. Scenarios~2.a and 2.b describe the data with less accuracy than Scenarios~1.a and 1.b (see $D/\rm ndf$ values in Table~\ref{tab:resultstable}). However, note that this spectral index allows us to expand the energy range where the fit is performed down to $5$~EeV. As in Scenarios~1.a and~1.b, the best-fit contribution from Vir~A is consistent with zero for both Scenarios~2.a and~2.b, indicating that the energy-spectrum fit alone does not require a significant contribution from this source. The contribution of background sources to the total flux is smaller than in Scenarios~1.a and 1.b due to the extended energy cutoff of DPLEC (boosted by $\Gamma_{\rm eff}^2$) and the extended fit range that provides additional limitations to $\eta_{\rm cont}$. While the predictions of Scenarios~1 qualitatively reproduce the observed trend toward a lighter composition at lower energies, the predicted mass evolution in Scenarios~2 is less steep and, as an \emph{a posteriori} comparison, is inconsistent with the Auger composition measurements below $\sim 20$~EeV.

Despite providing an acceptable fit to the energy spectrum, the predicted mass evolution in Scenario~2.b is flatter than indicated by the Auger composition measurements. Although the composition observables are not included in the fit, this \emph{a posteriori} comparison suggests that Scenario~2.b is not supported by the available composition data.

Figure~\ref{fig:fermi2.3all} shows the results for Scenario~3 which was introduced to assess the importance of the spectral index of the injected flux. In this Scenario, Vir~A emerges as a prominent source and the background sources contribution to the total flux becomes negligible. This is the only Scenario in which $\eta_{\rm For~A} < 1$, although it remains two orders of magnitude higher than $\eta_{\rm Cen~A}$ and one higher than $\eta_{\rm Vir~A}$. On the other hand, $\eta_{\rm Cen~A}$ for Cen~A continues consistent with the found in Scenarios 1 and 2. The predicted mass evolution exhibits the same excessive flatness observed in Scenarios~2.a and 2.b. 

\subsection{Secondary neutrinos}

\begin{figure*}[t]
    \begin{minipage}[b]{0.49\linewidth}
        \centering
        \includegraphics[width=\textwidth]{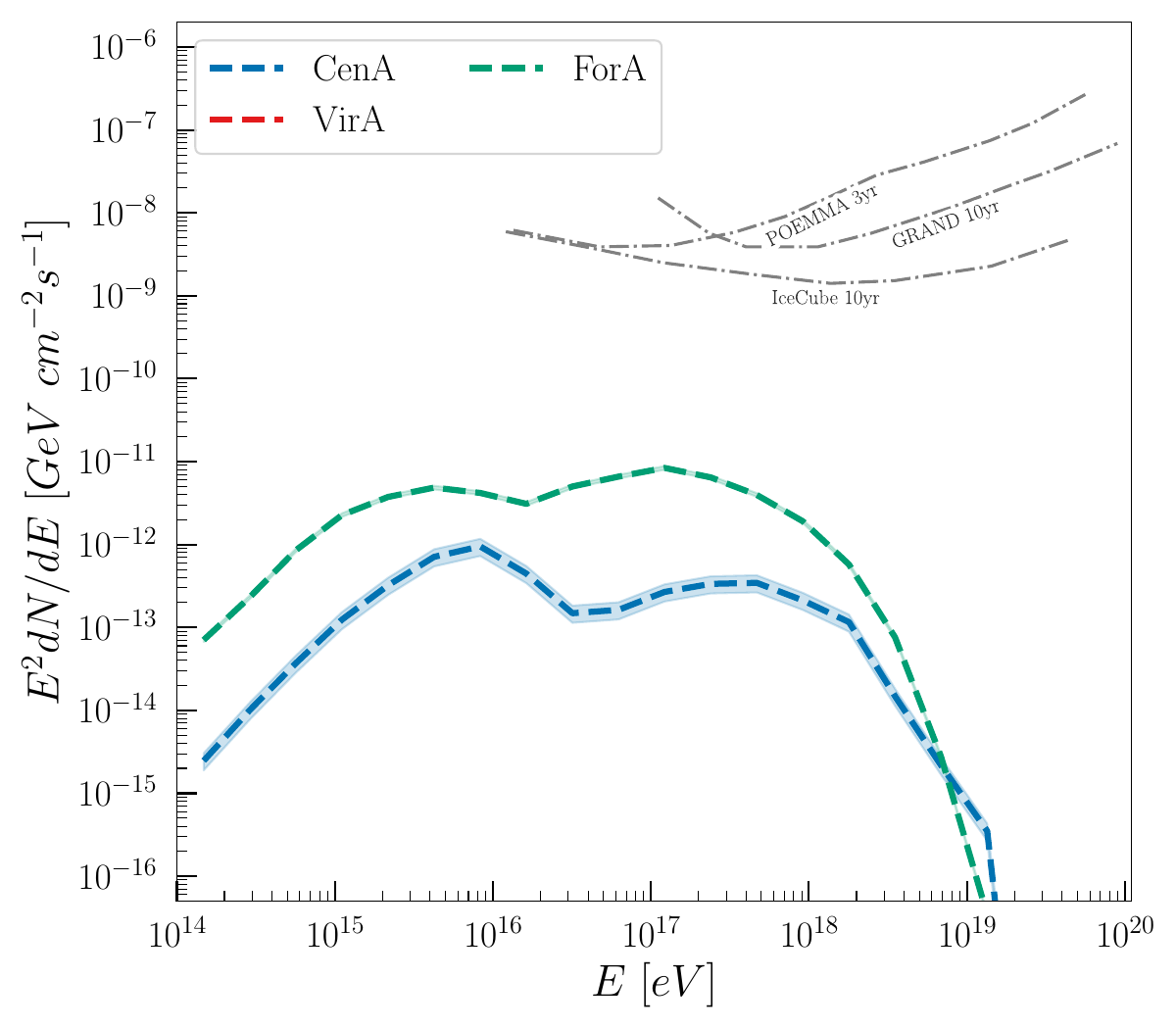}        
    \end{minipage}
    \hspace{0.5cm}
    \begin{minipage}[b]{0.49\linewidth}
        \centering
        \includegraphics[width=\textwidth]{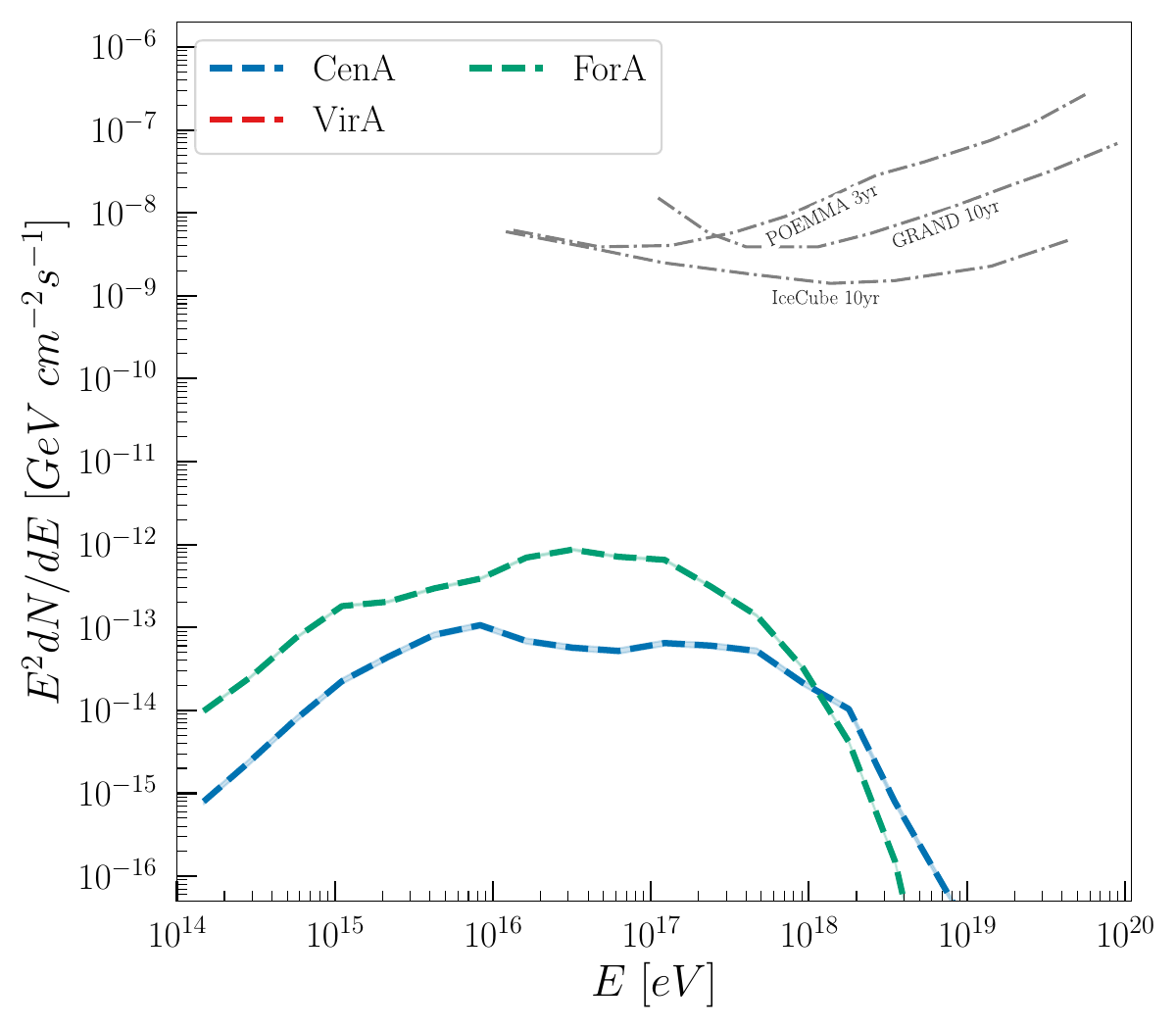}
    \end{minipage}
    \caption{Flux of secondary neutrino for Scenarios 1.a (\texttt{left}) and 2.a. (\texttt{right}) compared with the sensitivities of future experiments POEMMA~\cite{Venters_2020}, GRAND~\cite{grand} and IceCube-Gen2~\cite{icecube}. The uncertainties are shown as a shaded band.}
    \label{fig:local_neutrinos}
\end{figure*}

\begin{figure*}[t]
    \centering
    \includegraphics[width=0.49\linewidth]{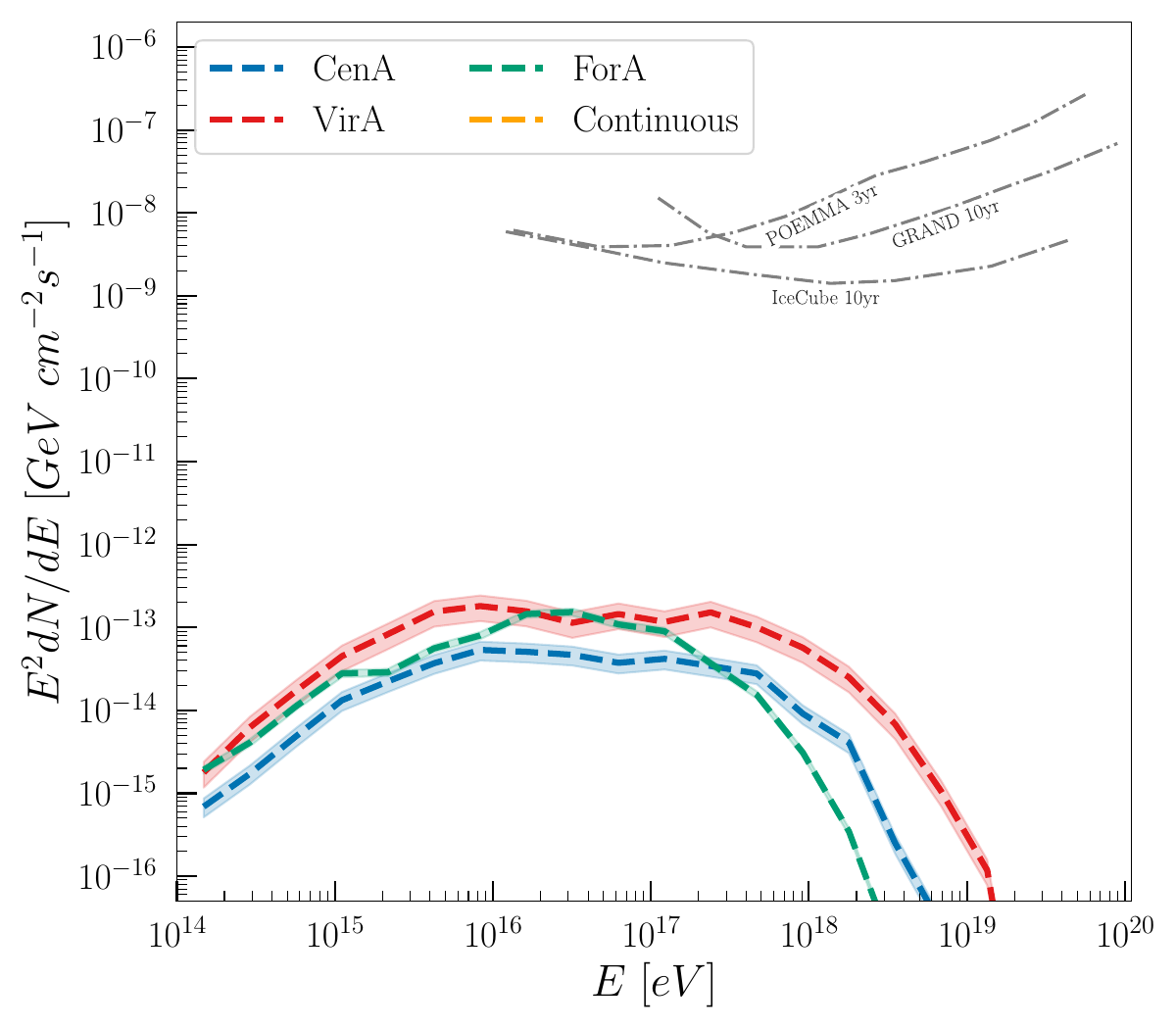}
    \caption{Flux of secondary neutrino for Scenario 3 compared with the sensitivities of future experiments POEMMA~\cite{Venters_2020}, GRAND~\cite{grand} and IceCube-Gen2~\cite{icecube}. The uncertainties are shown as a shaded band.}
    \label{fig:Neutrinos_3}
\end{figure*}

The neutrinos produced by the interactions of UHECRs with the photon backgrounds were calculated for each Scenario. Figure~\ref{fig:local_neutrinos} shows the flux of secondary neutrinos for Scenarios~1.a and 2.a and figure~\ref{fig:Neutrinos_3} for Scenario 3 in comparison to the sensitivities of future experiments POEMMA \cite{Venters_2020}, GRAND \cite{grand} and IceCube-Gen2 \cite{icecube}. The flux of secondary neutrinos from local sources is $3-4$ orders of magnitude below the sensitivities of future experiments. The choice of the energy spectrum model has a minor effect on the predicted neutrino flux. The production of secondary neutrinos is highly sensitive to the composition given that lighter nuclei produce more neutrinos~\cite{2005_hooper,2024_chakraborty}. As the fraction of protons in the \textit{Wolf-Rayet} composition is negligible, a very low flux of secondary neutrinos is produced.

\begin{figure*}[t]
    \begin{minipage}[b]{0.49\linewidth}
        \centering
        \includegraphics[width=\textwidth]{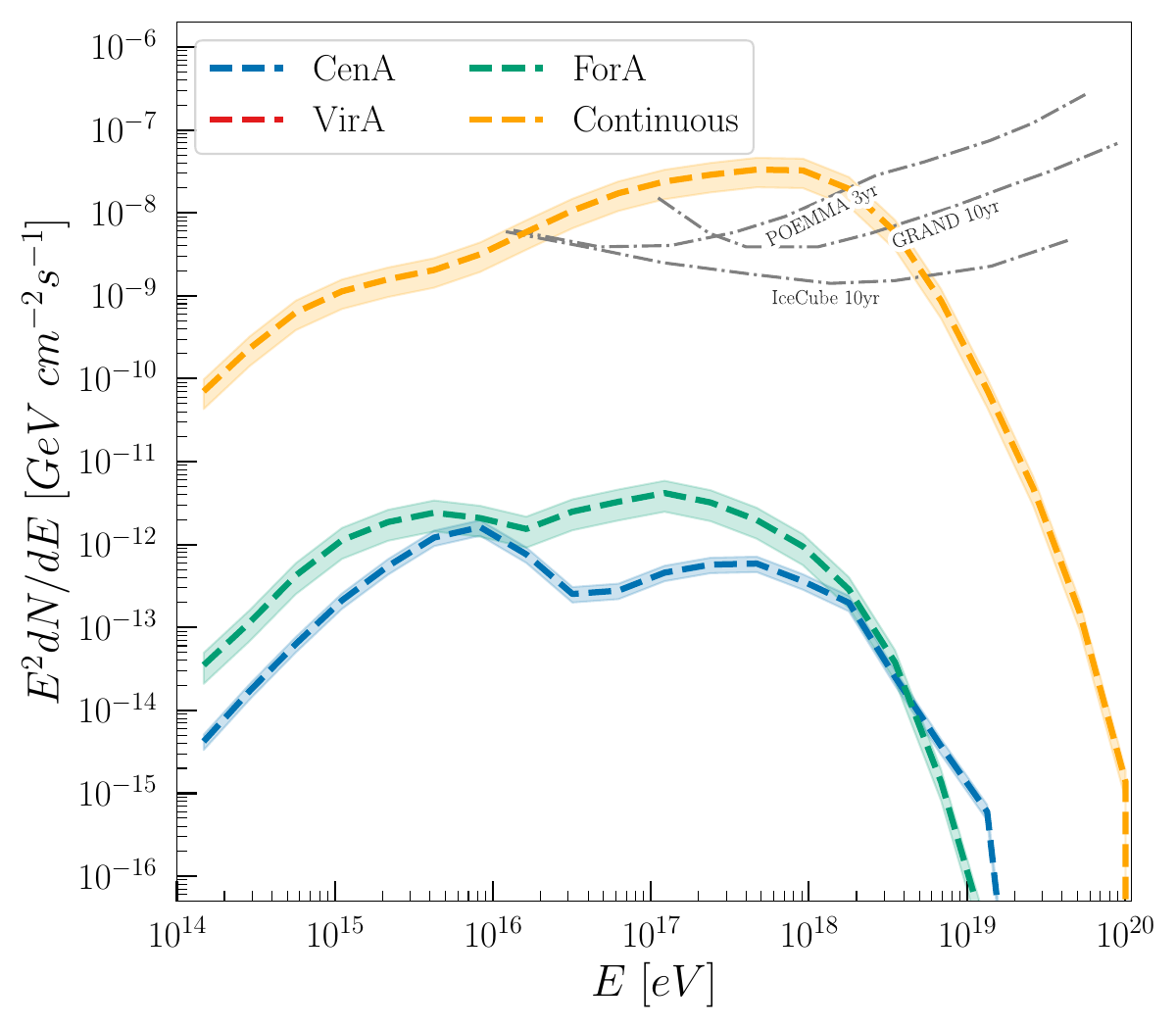}        
    \end{minipage}
    \hspace{0.5cm}
    \begin{minipage}[b]{0.49\linewidth}
        \centering
        \includegraphics[width=\textwidth]{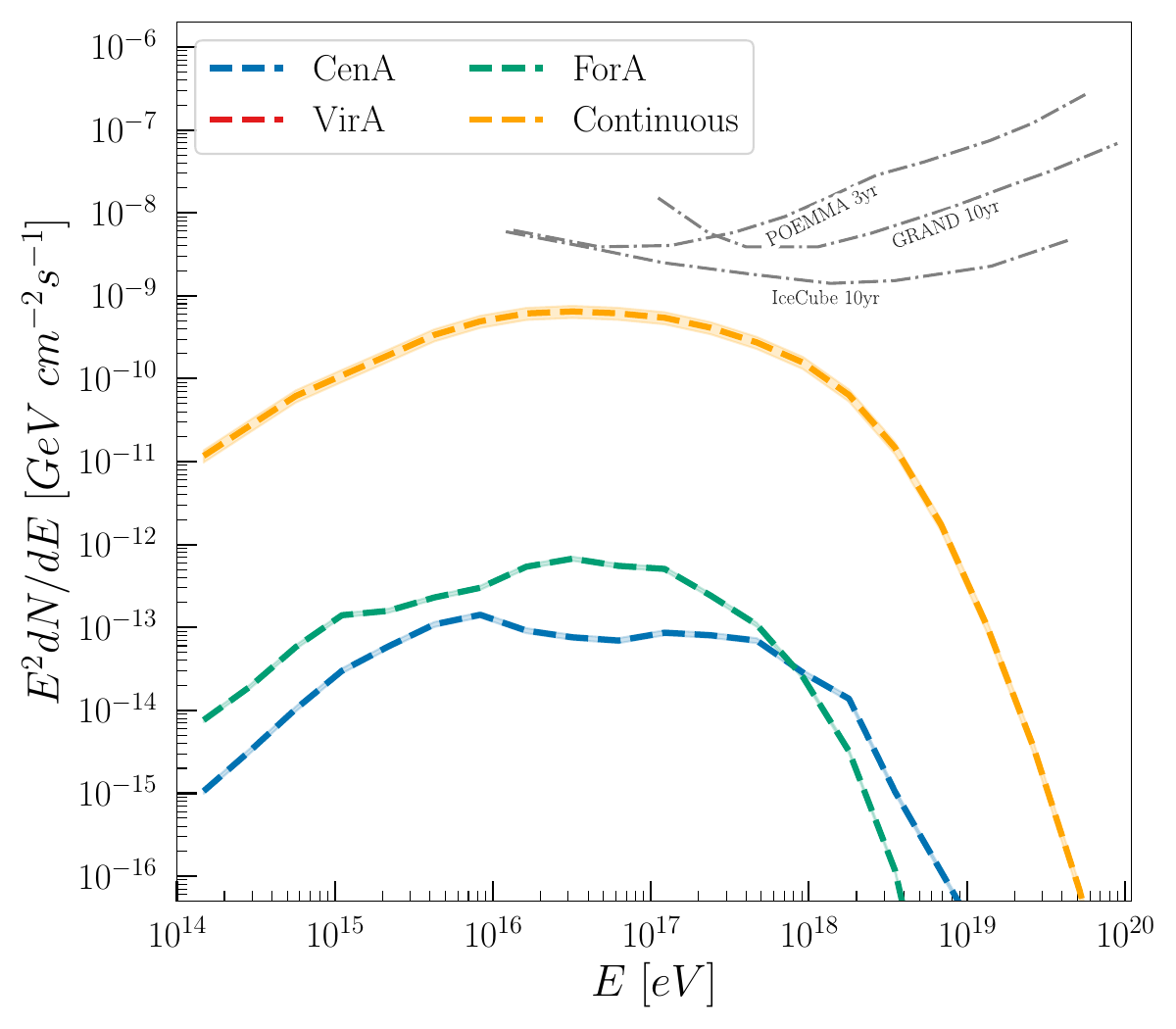}
    \end{minipage}
    \caption{Flux of secondary neutrino for Scenarios 1.b.~(\texttt{left}) and 2.b. (\texttt{right}) compared with the sensitivities of future experiments POEMMA \cite{Venters_2020}, GRAND \cite{grand} and IceCube-Gen2 \cite{icecube}. The uncertainties are shown as a shaded band.}
    \label{fig:local_neutrinos_2}
\end{figure*}

Figure~\ref{fig:local_neutrinos_2} shows the flux of secondary neutrinos for Scenarios~1.b and~2.b, in which background sources represent an important contribution to the flux measured at Earth. The neutrino yield depends on both the distance traveled by UHECRs and the number of contributing sources, both of which increase with distance. For the PLEC energy spectrum, the simulated neutrino flux exceeds the projected sensitivities of future experiments by a factor of approximately $\sim 10$, raising the exciting possibility of detecting UHECR-induced neutrinos from distant AGNs in the near future.

\section{Discussion}
\label{sec:discussion} 

 Table~\ref{tab:relativecontributions} presents the contribution of each source, for each Scenario, to the total flux measured by the Pierre Auger Observatory in the energy range over which the fits were performed. For the Scenarios considered, the total flux is largely dominated by local sources. The exact fraction of UHECRs originating from a given source, however, depends strongly on the assumed shape of the energy spectrum (PLEC or DPLEC), with the contribution from background sources remaining at the level of $\lesssim 20\%$.

The relative source fractions reported in Table~\ref{tab:relativecontributions} should be interpreted within the assumptions of the present propagation model. In particular, the neglect of extragalactic and Galactic magnetic fields implies that the fitted fractions correspond to effective contributions at Earth in the absence of magnetic deflections. In a more realistic three-dimensional scenario, magnetic deflections, time delays, and magnetic lensing may either suppress or enhance the apparent flux from individual nearby sources. Consequently, the inferred source fractions should not be interpreted as precise measurements of the intrinsic UHECR luminosity of each radio galaxy.

\begin{table*}[t]

    \caption{Relative contribution of each source to the total flux measured by the Pierre Auger Observatory within the fit range.}
     \label{tab:relativecontributions}

    \begin{tabular*}{\textwidth}{@{\extracolsep\fill}cccccc}
    \hline
    Scenario & Energy Range ($\rm EeV$) & Cen~A & Vir~A & For~A & Continuous \\ \hline
    1.a    & $20-200 $ & $30\pm7\%$  & $0\pm20\%$  & $66\pm3\%$   &  -                 \\
    1.b    & $20-200$  & $51\pm11\%$ & $0\pm23\%$  & $33\pm13\%$  &  $15\pm6\%$      \\
    2.a    & $5-200 $  & $46\pm3\%$  & $0\pm1\%$   & $46\pm1\%$   & -                  \\
    2.b    & $5-200 $  & $62\pm4\%$   & $0\pm19\%$  & $35\pm2\%$   & $1.6\pm0.3\%$    \\
    3    & $5-200 $    & $54\pm14\%$ & $31\pm11\%$ & $12\pm1\%$   &  $0.0\pm0.1\%$ \\ \hline
    \end{tabular*}
    
    \end{table*}

\textbf{Cen~A dominance.} Cen~A exhibits a significant contribution across all Scenarios, with typical values of $\eta \sim 10^{-3}$. Due to its proximity and current jet power, it consistently emerges as the dominant source of the UHECR flux in the Scenarios considered. This result is qualitatively interesting given the reported excess of events in the Cen A region, although no anisotropy modeling is included in the present work. 

The fitted values of $\eta$ obtained for Cen~A (see Table~\ref{tab:resultstable}) are broadly compatible with previous estimates based on physically motivated models of particle acceleration in AGN jets, such as those of Matthews et al.~\cite{Matthews_2018_backflow}. However, as discussed above, $\eta$ should be interpreted as an effective normalization parameter rather than as a direct measurement of the physical acceleration efficiency. Consequently, the comparison is meaningful only at the order-of-magnitude level, while the relative trends between scenarios are more informative than the absolute numerical values. If instead a relativistic shock spectrum with $s = 2.3$ is assumed, the resulting values of $\eta$ are reduced by less than a factor of two, still remaining within the same order of magnitude.

\textbf{For~A past.} A good description of the energy spectrum measured by the Pierre Auger Observatory is obtained for $\eta_{\rm For~A} > 1$. Aside from the discussed uncertainties in the jet power, these values are only achievable if For~A presents a powerful past. In fact, For~A exhibits some signs of higher activity in the past~\cite{Maccagni_2020}, and the necessity of it to explain the arrival directions of UHECRs was first pointed out by Matthews et al (2018)~\cite{Matthews_2018}. For~A direction can be correlated with a region in which an excess of events above $40 \rm EeV$ was found in the Pierre Auger Observatory data~\cite{Abreu_2022, de_Oliveira_2023}. As the propagation times of UHECR are not equivalent to those for radio waves used as proxies for jet power measurements, particles could have undergone acceleration in a more extreme environment. To match the acceleration efficiency $\eta$ found for other sources, For~A should have been between $2$ and $3$ orders of magnitude more powerful in the past. However, the maximum energy can also be connected to the jet power. This increase in $Q_{\rm jet}$ will shift $E_{\rm break}$ by factors $\sim15-50$ (Eq.~\ref{eq:Ebreak_app}). An alternative Scenario could be the magnetic amplification of the flux from For~A, as obtained in de Oliveira \& de Souza~\cite{de_Oliveira_2022}, however, only modest amplification factors ($\sim2$) were observed. One possible explanation is related to time-delays imposed by the propagation of UHECR in the giant radio lobes~\cite{Matthews_2018, deOliveira_2025_history}. For~A's lobes extend to $122-170 \rm kpc$, with an average magnetic field $\sim 3 \rm \mu G$ \cite{Maccagni_2020}. Comparing the results from \cite{deOliveira_2025_history} for Cen~A ($\sim 200 \rm~kpc$, $\sim 1 \rm~\mu G$), it is useful to get a qualitative picture. In this case, if For~A presented a powerful past, UHECR with energies $\lesssim 30$~EeV may remain confined in the lobes for longer $\sim 0.1-1~\rm Myr$, while more energetic $\sim 10^{20}$~eV escape more quickly. More definitive conclusions require a deeper study of the confinements of UHECR in the lobes of For~A.
   
\textbf{Vir~A absence.} In Scenarios~1 and~2, the best-fit contribution from Vir~A is consistent with zero within the statistical uncertainties, whereas Scenario~3 yields a substantially larger contribution from this source, amounting to $(31 \pm 10)\%$. This increase is driven by the change in the spectral index $s$ of the injected energy spectrum. Larger values of $s$ reduce the relative contribution of more distant sources, thereby forcing the fit to reproduce the observed flux predominantly through nearby sources, namely Cen~A and Vir~A.

Given its high jet power, Vir~A remains a plausible UHECR accelerator candidate. Owing to its relative proximity, Vir~A has long been advocated as a prominent contributor to the UHECR sky if active galactic nuclei are the dominant sources of these particles. For this reason, the persistent absence of an excess of UHECR events in the direction of Vir~A, as reported by all UHECR experiments to date, has remained a puzzling feature for many years~\cite{2018kobzar}.

In this context, the fact that the spectrum-only fit does not require a significant contribution from Vir~A may indicate the presence of a suppression mechanism acting on the observable UHECR flux from this source. However, the current fit alone is not sufficient to establish such a mechanism. One possible explanation involves the structure of the extragalactic magnetic field (EGMF), which may significantly suppress or deflect the UHECR flux from Vir~A. The galactic magnetic field (GMF) can also de-magnify its flux, as shown by Bister et. al~\cite{Bister_2024}, with greater prominence for rigidities $\sim 1 EV$. An alternative hypothesis is related to the internal structure of the source: cosmic-ray diffusion within the giant radio lobes of Vir~A may be sufficiently slow to strongly limit the observable flux. Another plausible Scenario is that Vir~A was less powerful in the past, at the epoch when UHECR acceleration would have occurred.

As discussed by Matthews et al.~\cite{Matthews_2018}, the radio lobes of Vir~A appear to be predominantly inflated by the current jet activity~\cite{Owen_2000}, in contrast to Cen~A and For~A, which show indications of a more violent past. In such a case, UHECRs originating from Vir~A may not yet have had sufficient time to reach the Earth.

The present treatment assumes stationary source emission. However, finite source activity times and propagation delays may significantly affect the observable UHECR flux from nearby radio galaxies, particularly in the presence of structured magnetic fields. Such effects could alter both the spectral normalization and the apparent contribution of individual sources, especially for Virgo A and Fornax A. A time-dependent treatment is deferred to future work.

The predicted cosmogenic neutrino flux is considerably more model dependent than the UHECR energy spectrum. The dominant uncertainty is the contribution from the distant radio-galaxy population, since UHECRs from nearby sources undergo relatively few interactions during propagation, whereas particles originating at cosmological distances have a much higher probability of interacting with the cosmic microwave and extragalactic background radiation fields. Consequently, the neutrino flux is particularly sensitive to the adopted radio luminosity function and the cosmological evolution of the source population. In addition, the injected mass composition plays a major role: the Wolf-Rayet composition adopted in this work contains a negligible proton fraction, substantially suppressing neutrino production compared to lighter compositions. The neutrino predictions presented here should therefore be interpreted within the context of these modeling assumptions.

\section{Conclusion}
\label{sec:conclusion}

In this work, we presented a new, physically grounded framework to investigate the origin of ultra-high-energy cosmic rays (UHECRs) in nearby radio galaxies. By combining acceleration spectra derived from detailed modeling of particle acceleration in relativistic jets with numerical simulations of extragalactic propagation, we performed a source-resolved analysis of the three nearest FR-I radio galaxies—Centaurus~A, Virgo~A, and Fornax~A—while consistently accounting for the cumulative contribution of the more distant radio-galaxy population.

A central result of this study is that the UHECR energy spectrum measured by the Pierre Auger Observatory can be reproduced with a minimal set of physically motivated parameters. In the scenarios that best describe the data, the observed flux at the highest energies can be reproduced predominantly by nearby sources within the assumptions of the present propagation model. The inferred relative contributions of individual radio galaxies should be regarded as effective source fractions and may be modified by magnetic deflections, time delays, and magnetic lensing in more realistic three-dimensional propagation scenarios. The required acceleration efficiencies correspond to only a small fraction of the mechanical jet power, typically in the range $\eta \sim 10^{-3}$–$10^{-2}$, which is fully compatible with current theoretical expectations for particle acceleration in AGN jets.

Among the individual sources, Centaurus~A consistently emerges as a major contributor across all viable Scenarios, in line with its proximity and jet power. Fornax~A can also play a significant role, although the inferred efficiencies suggest that past activity or time-delay effects associated with particle confinement in extended radio lobes may be important. In contrast, the spectrum-only fit does not require a significant contribution from Virgo~A in most scenarios, although Scenario~3 favors a substantially larger contribution.

The framework developed here also yields non-trivial predictions for the mass composition at Earth and for the associated flux of secondary neutrinos. While the model is fitted exclusively to the energy spectrum, the resulting composition trends are broadly consistent with observational constraints at the highest energies, and the predicted neutrino fluxes provide an additional, independent avenue for testing the scenario with future experiments.

Overall, this study shows that nearby radio galaxies remain plausible contributors to the highest-energy component of the observed UHECR flux within a physically motivated source-resolved framework. By moving beyond simplified injection assumptions and homogeneous source populations, our results highlight the importance of source-specific modeling in UHECR studies and establish a establish a physically motivated framework that can be further refined and tested with improved composition measurements, anisotropy studies, and next-generation multi-messenger observations.

The present study adopts one-dimensional propagation without magnetic fields in order to isolate the impact of source energetics, acceleration spectra, and nuclear composition on the observable energy spectrum. This approximation is expected to be more reliable at the highest rigidities, where magnetic deflections become smaller. Future developments including magnetic-field effects, anisotropy observables, and composition-constrained fitting will be essential to further test the robustness of these scenarios.

\backmatter
\bmhead{Acknowledgements}
 The authors acknowledge FAPESP and CNPq for financial support and the National Laboratory for Scientific Computing (LNCC\slash MCTI,  Brazil) for providing HPC resources for the SDumont supercomputer (http://sdumont.lncc.br).

 This version of the article has been accepted for publication after peer review, but is not the Version of Record and does not reflect post-acceptance improvements, or any corrections. The Version of Record is available online at: https://doi.org/10.1140/epjc/s10052-026-16558-3

\section*{Declarations}

\textbf{Funding:}
The authors are supported by the S\~{a}o Paulo Research Foundation (FAPESP) through grant numbers 2021\slash01089-1, 2020\slash15453-4, 2024\slash22747-5, 2024\slash22722-2, 2024\slash22721-6, 2025\slash03325-5 and 2019\slash10151-2. VdS is supported by CNPq through grant number 308837\slash2023-1. 

\textbf{Competing interests:}
The authors declare that they have no known competing financial interests or personal relationships that could have appeared to influence the work reported in this paper.

\textbf{Employment:}
Not applicable.

\textbf{Ethics approval and consent to participate:}
Not applicable.

\textbf{Consent for publication:}
Not applicable.

\textbf{Data and code availability:}
The datasets and code generated during the current study are available from the corresponding author on reasonable request.

\begin{appendices}

\section{Simulation weighting}
\label{appendix:normalization}

The simulations described in ~\ref{sec:propagation} were weighted to correct for inherent simulation biases and ensure the physical problem is treated correctly. For individual sources, each particle detected by the observer was assigned a weight that accounts for the emitted spectrum by the source and the source–Earth distance $D$. Accordingly, for each particle $i$ detected by the observer we define
\begin{equation}
w^i = \frac{1}{4\pi D^2}\frac{dN_j}{dEdt}\frac{E_0^i \ln(10^{3})}{N_j},
\end{equation}
where $N_j$ is the total number of injected particles of nuclear species $j$. The factor $E_0 \ln(10^{3})$ results from the normalization of the injected spectrum, given by
\begin{align}
\left[\frac{E_0^{-1}}{\int_{E_{\text{min}}}^{E_{\text{max}}} dE_0 E_0^{-1}}\right]^{-1}
= E_0 \ln\left(\frac{E_{\text{max}}}{E_{\text{min}}}\right) \nonumber \\
=  E_0 \ln(10^3)
\end{align}

For continuous sources, the emission rate is obtained from the emissivity described in Section~\ref{sec:continuous_sources_model} and each shell $k$ must be normalized by the injected particle density number $\rho^{-1} = (D_{\text{max}} - D_{\text{min}})/N_k$ in simulation. Hence, the weight is 
\begin{equation}
w^i = \frac{1}{4}\frac{dN_j}{dVdtdE} E^i_0\ln(10^3)\frac{D_{\text{max}} - D_{\text{min}}}{N_k},
\end{equation}
where the factor $1/4$ accounts for the transition from the 1D to the 3D treatment.
\section{Relation between jet power and break energy} 
\label{appendix:break_energy}

\begin{figure}[htbp]
    \centering
    \includegraphics[width=\linewidth]{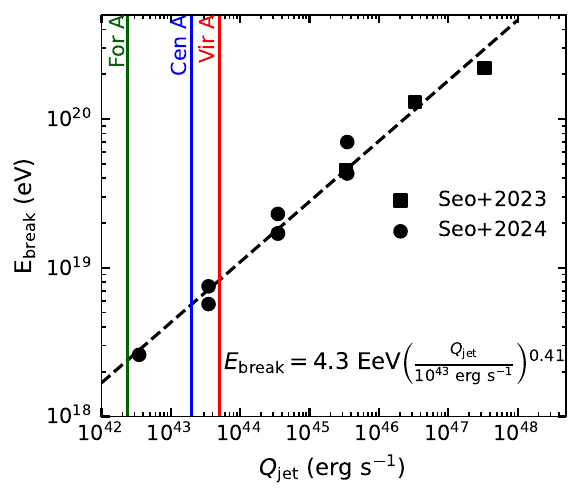}
    \caption{Relation of $E_{\rm break}$ and $Q_{\rm jet}$ for the data of Seo et al. (2023)~\cite{Seo_2023} (squares) Seo et al. (2024)~\cite{Seo_2024} (circles). The best fit is shown in a black dashed line. The estimated jet power for the nearby sources Cen~A (blue), Vir~A (red), and For~A (green) is also shown.}
    \label{fig:Ebreak}
\end{figure}

Seo et al. ~\cite{Seo_2023,Seo_2024} obtained $E_{\rm break}$ for several conditions, such as kinetic power, radius, and Lorentz factor of the jet. Based on the Hillas-Lovelace limit~\cite{1976_Lovelace, 1984_Hillas}, we expect a dominance dependence on the jet kinetic power, given by $E_{\rm break} \propto Q_{\rm jet}^{1/2}$. Figure~\ref{fig:Ebreak} shows $E_{\rm break}$ as a function of $Q_{\rm jet}$ for the results in references~\cite{Seo_2023,Seo_2024}. We estimate a general relation between the break energy and the kinetic power of the jet as
\begin{equation} \label{eq:Ebreak_app}
    E_{\rm break} = 4.3~{\rm EeV} \left( \frac{Q_{\rm jet}}{10^{43}~\rm {erg~s}^{-1}} \right)^{0.41}.
\end{equation}

Note that the obtained dependence of $E_{\rm break}$ with $Q_{\rm jet}$ is weaker than predicted by the ideal Hillas-Lovelace limit, indicating a less efficient acceleration.

\end{appendices}

\bibliography{biblio}

\end{document}